\documentclass[
reprint,
superscriptaddress,
amsmath,amssymb,
aps,
pra,
]{revtex4-1}

\usepackage{graphicx}
\usepackage{dcolumn}
\usepackage{upgreek}
\usepackage{mathrsfs}
\usepackage[breaklinks=true,colorlinks=true,linkcolor=blue,urlcolor=blue,citecolor=blue]{hyperref}
\allowdisplaybreaks[4]
\begin{document}

\preprint{APS/123-QED}

\title{Manipulating photonic quantum states with long-range interactions}
\author{Fan Yang}
\affiliation{State Key Laboratory of Low Dimensional Quantum Physics, Department of Physics, Tsinghua University, Beijing 100084, China}
\author{Yong-Chun Liu}
\email{ycliu@mail.tsinghua.edu.cn}
\affiliation{State Key Laboratory of Low Dimensional Quantum Physics, Department of Physics, Tsinghua University, Beijing 100084, China}
\affiliation{Collaborative Innovation Center of Quantum Matter, Beijing, China}
\author{Li You}
\email{lyou@mail.tsinghua.edu.cn}
\affiliation{State Key Laboratory of Low Dimensional Quantum Physics, Department of Physics, Tsinghua University, Beijing 100084, China}
\affiliation{Collaborative Innovation Center of Quantum Matter, Beijing, China}


\begin{abstract}
We present a scheme for coherently manipulating quantum states of photons by incorporating multiple photonic modes in a system with long-range interactions. The presence of nonlocal photon-photon interactions destroys the energy or momentum matching conditions between distinct propagating polaritons, and consequently gives rise to blockaded effective coupling between the corresponding polaritons. Such a blockade mechanism protects the system from interaction-induced dissipations and enables highly tunable few-photon nonlinearities. Taking Rydberg atomic ensemble as an example, we illustrate several intriguing phenomena based on the proposed scheme, e.g., the deterministic generation of entangled photon pairs, the nonlinear beam splitting, as well as the establishment of a tunable dressed interaction between individual photons.
\end{abstract}

\maketitle
\section{Introduction}\label{sec:sec1}
Strong interactions between light quanta or photons can provide effective control of their quantum states \cite{chang2014quantum}. Significant progresses have been made towards realizing strong photonic nonlinearities \cite{chang2014quantum,murray2016chapter,firstenberg2016nonlinear}, which cultivated the development of research fields ranging from quantum information processing \cite{friedler2005long,kimble2008quantum,o2009photonic} to simulating many-body physics \cite{chang2008crystallization,otterbach2013wigner,gullans2016effective,murray2018photon}. One feasible realization employs strong coupling between highly localized single photons and a single quantum emitter to generate local Kerr nonlinearity \cite{birnbaum2005photon,thompson2013coupling,goban2014atom,reiserer2015cavity}. Alternatively, effective long-range interactions can be realized by coupling propagating photons to atomic vapor ensembles \cite{murray2016chapter,firstenberg2016nonlinear} or atomic arrays in waveguide QED systems \cite{douglas2016photon,shahmoon2016highly}.

A widely discussed scheme for strong and nonlocal optical nonlinearity is based on electromagnetically induced transparency (EIT) inside a medium of Rydberg atoms \cite{gorshkov2011photon,maghrebi2015coulomb,peyronel2012quantum,firstenberg2013attractive,baur2014single,gorniaczyk2014single}. The nonlinearity in this system arises from EIT blockade where a multi-photon input destroys the EIT condition \cite{gorshkov2011photon}, and gives rise to a variety of desirable features \cite{peyronel2012quantum,firstenberg2013attractive,baur2014single,gorniaczyk2014single,tresp2016single,distante2016storage}. Most of the earlier studies involve one type of photonic mode, inevitably limiting the scopes and capabilities of photonic information processing. In addition, unavoidable dissipation associated with EIT blockade can cause spatial decoherence to the lone phontic mode, which erects a performance bottleneck for several key applications \cite{gorshkov2013dissipative,murray2016many}.

We show in this work that by incorporating additional photonic modes and long-range photon-photon interactions, flexible and coherent manipulation of photonic quantum states can be carried out with high fidelities. In contrast to the pioneering proposal of Ref.~\cite{murray2017coherent} based on interaction caused polariton mode conversion, our scheme makes use of mode conversion blockade from interaction caused mismatches. Photonic polaritons satisfying energy and momentum matching conditions can effectively couple to one another during propagation \cite{yariv2007photonics}. The presence of strong and long-range photonic interactions destroys the matching conditions, and consequently causes multi-excitations of the nonlinear modes to be blockaded. The subsequent evolution thus restricts the system to an interaction-free subspace, immune to interaction-induced dissipations. Such a blockade mechanism enables the flexible control of few-photon nonlinearities, and facilitates a variety of quantum information protocols, as discussed below.

\section{A generic model}\label{sec:sec2}
In this section, we build up a generic model to describe the basic blockade mechanism, which is not restricted to specific realizations. To illustrate our idea, we first consider two counterpropagating photons, Appendix \ref{app:appA} details the case for copropagating. Four photonic polariton fields are involved, of which the two forward (backward) propagating ones are described by their slowly-varing operators $\hat{\Psi}_{a_+}(z)$ and $\hat{\Psi}_{b_+}(z)$ [$\hat{\Psi}_{a_-}(z)$ and $\hat{\Psi}_{b_-}(z)$], with $\mu_{\pm}$ ($\mu=a,b$) denoting specific polariton modes. Modes $a_\pm$ and $b_\pm$ are linearly coupled. Photons in modes $a_+$ and $a_-$ interact nonlocally, while photons in all other modes are assumed noninteracting. The Hamiltonian of our model takes the form $\hat{H}=\hat{H}_0^+ + \hat{H}_0^- + \hat{H}_\mathrm{c}^+ + \hat{H}_\mathrm{c}^- +\hat{H}_\mathrm{int}$, where
\begin{align}
\label{eq:eq1}
\hat{H}_0^\pm =& -\sum_{\mu=a,b}\pm iv_{\pm}\int dz\hat{\Psi}_{\mu_\pm}^\dagger(z) \partial_z\hat{\Psi}_{\mu_\pm}(z),\\
\label{eq:eq2}
\hat{H}_\mathrm{c}^\pm =& \int dz g_\pm\hat{\Psi}_{a_\pm}^\dagger(z)\hat{\Psi}_{b_\pm}(z)e^{i\Delta k_\pm z}+\mathrm{H.c.},\\
\label{eq:eq3}
\hat{H}_\mathrm{int} =& \int dz dz^\prime \mathcal{V}(r)\hat{\Psi}_{a_+}^\dagger(z)\hat{\Psi}_{a_-}^\dagger(z^\prime)\hat{\Psi}_{a_-}(z^\prime)\hat{\Psi}_{a_+}(z),
\end{align}
denote the equivalent photon kinetic energy with group velocity $v_\pm$, the beam-splitting coupling with strength $g_\pm$ and detuning $\Delta k_\pm$, and the nonlocal two-body interaction via potential $\mathcal{V}(r)$ with $r=z-z^\prime$, respectively.

The evolution of the two-photon quantum state $|\psi(t)\rangle$ is described by the two-photon wavefunction $\Psi_{\mu_+\nu_-}(z_1,z_2,t)=\langle0|\hat{\Psi}_{\mu_+}(z_1)\hat{\Psi}_{\nu_-}(z_2)|\psi(t)\rangle$. Defining center-of-mass and relative coordinates $R=(v_{-}z_1+v_{+}z_2)/(v_{+}+v_{-})$ and $\rho=z_1-(v_{-}/v_{+})z_2$, and moving into the retarded time frame with $\tau=t-\rho/v_\mathrm{eff}$, $\xi=t$, and $v_\mathrm{eff}=(v_+^2+v_-^2)/v_+$, one obtains $i\partial_\xi\psi=\mathcal{H}\psi$, where
\begin{equation}
\label{eq:eq4}
\psi=\left(\tilde{\Psi}_{a_+a_-},\tilde{\Psi}_{a_+b_-},\tilde{\Psi}_{b_+a_-},\tilde{\Psi}_{b_+b_-}\right)^T
\end{equation}
is the two-photon wavefunction in the rotating frame, and the effective Hamiltonian becomes
\begin{equation}
\label{eq:eq5}
\mathcal{H}=\begin{pmatrix}
v_{+}\Delta k_+-v_{-}\Delta k_-+\mathcal{V}&g_-&g_+&0\\
g_-&v_{+}\Delta k_+&0&g_+\\
g_+&0&-v_{-}\Delta k_-&g_-\\
0&g_+&g_-&0
\end{pmatrix}.
\end{equation}
The potential $\mathcal{V}(r)$ is in general a function of both $\tau$ and $\xi$. However, if the long-range interaction is strong enough to meet the condition $|\mathcal{V}|\gg g_\pm, v_\pm|\Delta k_\pm|$ for all $r$ within the range of the wavefunction, the conversion of $\tilde{\Psi}_{a_+b_-}$ and $\tilde{\Psi}_{b_+a_-}$ to $\tilde{\Psi}_{a_+a_-}$ will both be blockaded due to the large energy detuning (or momentum mismatch), irrespective of the exact value or form of $\mathcal{V}$. This coupling blockade for propagating photons arises with long-range interaction, which forbids multi-excitations of the nonlocal nonlinear mode, with its desirable features reminiscent of the widely discussed Rydberg blockade \cite{lukin2001dipole,urban2009observation,gaetan2009observation}.

\section{Blockade mechanism in Rydberg EIT systems}\label{sec:sec3}

Various approaches for realizing strong nonlocal interaction between individual photons are discussed \cite{douglas2016photon,shahmoon2016highly,gorshkov2011photon}. In this study, we adopt the Rydberg EIT implementation \cite{gorshkov2011photon} with mode $a_\pm$ the dark state polariton (DSP) $\hat{\Psi}_{a_\pm}(z)=\cos\theta\hat{\mathcal{E}}_{a_\pm}(z) -\sin\theta\hat{\mathcal{S}}_{a_\pm}(z)$, a superposition of electromagnetic field $\hat{\mathcal{E}}_{a_\pm}(z)$ and Rydberg spin-wave field $\hat{\mathcal{S}}_{a_\pm}(z)$, with $\tan\theta=g_p/\Omega$, $g_p$ the collective atom-photon coupling strength, and $\Omega$ the Rabi frequency of the control field \cite{fleischhauer2000dark}. The van der Waals interaction $V(r)=C_6/r^6$ between the Rydberg spin-wave components $\hat{\mathcal{S}}_{a_\pm}(z)$ can induce strong and long-range interactions between DSP modes $a_+$ and $a_-$ \cite{murray2016chapter}.

\begin{figure}
\centering
\includegraphics[width=\linewidth]{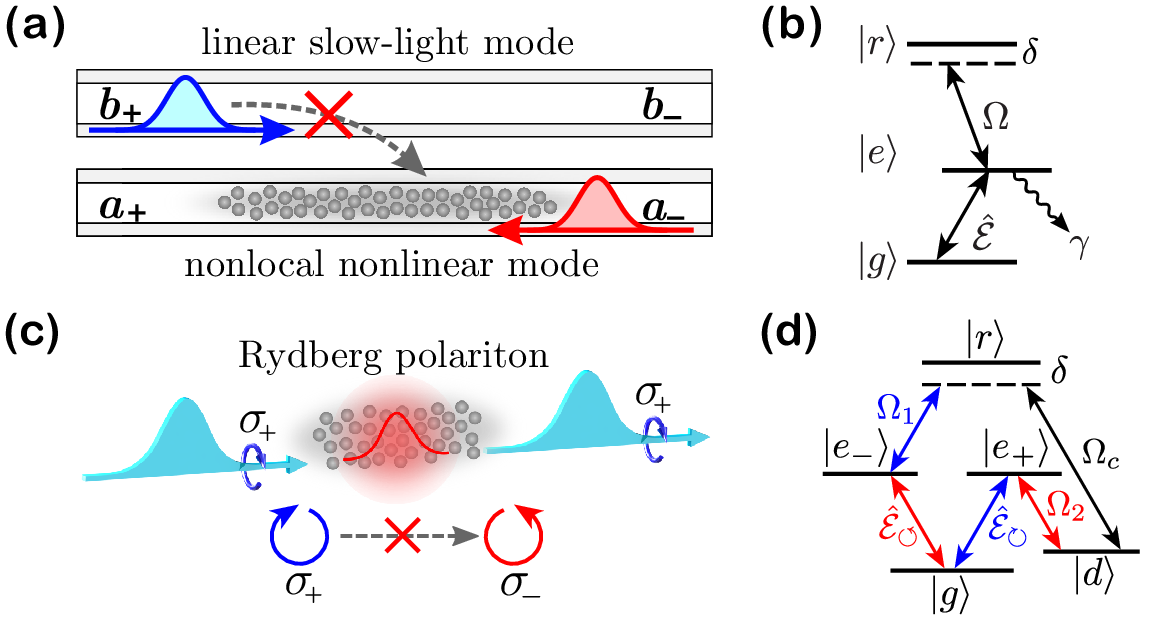}
\caption{(a) A schematic for the coupling blockade for counterpropagating photons in a waveguide-like system. (b) Rydberg EIT level diagram, in which the ground state $|g\rangle$, the excited state $|e\rangle$ (linewidth $2\gamma$), and the Rydberg state $|r\rangle$ are coupled respectively by a quantum probe field $\hat{\mathcal{E}}$ and a classical control field $\Omega$, with a two-photon detuning $\delta$. (c) Spin-flip blockade of the propagating photons and (d) the associated atomic level structure.}
\label{fig:fig1}
\end{figure}

We present two practical implementations for constructing the linear coupling Hamiltonian. (i) In the photonic coupling scheme shown in Figs.~\ref{fig:fig1}(a) and \ref{fig:fig1}(b), we take $\hat{\Psi}_{b_\pm}(z)=\hat{\mathcal{E}}_{b_\pm}(z)$ to be an electromagnetic field of a linear slow-light mode and $\hat{\Psi}_{a_\pm}(z)$ to be a nonlinear DSP that for example can be produced in a hollow-core waveguide filled with Rydberg atoms \cite{christensen2008trapping,bajcsy2009efficient,vetsch2010optical,shahmoon2011strongly,langbecker2017rydberg}. The beam-splitting coupling $g_\pm\sec\theta\hat{\mathcal{E}}_{a_\pm}^\dagger(z)\hat{\mathcal{E}}_{b_\pm}(z)+\mathrm{H.c.}$ can be obtained from overlapping their evanescent fields. (ii) For the atomic coupling scheme shown in Figs.~\ref{fig:fig1}(c) and \ref{fig:fig1}(d), $\hat{\Psi}_{b_\pm}(z)$ represents a linear DSP composed of $\sigma_+$ photonic field $\hat{\mathcal{E}}_{\circlearrowright_\pm}(z)$ and ground state spin wave field $\hat{\mathcal{S}}_{b_\pm}(z)$, and $\hat{\Psi}_{a_\pm}(z)$ denotes a nonlinear DSP as a superposition of $\sigma_-$ photonic field $\hat{\mathcal{E}}_{\circlearrowleft_\pm}(z)$ and Rydberg spin wave $\hat{\mathcal{S}}_{a_\pm}(z)$. The linear coupling $g_\pm\csc^2\theta\hat{\mathcal{S}}_{a_\pm}^\dagger(z)\hat{\mathcal{S}}_{b_\pm}(z) +\mathrm{H.c.}$ can be controlled by the external field $\Omega_c$. In the slow-light regime, influences of coupling to bright state polaritons (BSPs) in the above two schemes are negligible. For simplicity, we assume that mode $a_\pm$ and $b_\pm$ propagate with matched group velocities, while a small mismatch does not significantly influence the coupling between them for short operation time (see Appendix \ref{app:appD}).

In the conventional EIT blockade \cite{gorshkov2011photon}, a dispersive photonic interaction can be established by using a large single-photon detuning $\Delta$, which, however, results in a large group velocity dispersion and can limit the performance of related applications \cite{he2014two}. For this reason, we take $\Delta=0$ in this work, where the induced interaction between nonlinear DSPs is highly dissipative. With the proposed blockade mechanism, we can transform such a dissipative interaction into a dispersive one, and achieve coherent manipulations of photonic quantum state.

To clarify the blockade mechanism, we take the backward photon to be a stored gate photon ($g_-,v_-=0$). In the Rydberg EIT implementation, this can be achieved by taking $\hat{\Psi}_{a_-}(z)=\hat{\mathcal{S}}_{-}(z)$ to be a stored spin wave of an auxiliary Rydberg state. The parameters for the forward target photon are $g_+=g$, $v_+=v$, and $\Delta k_+=0$.

For photonic coupling scheme, the spatial modulation of $g$ is considered, i.e., $g=g(z)$. The corresponding dynamics can be described by the two-photon wavefunction $E_aS=\langle 0|\hat{\mathcal{E}}_{a_+}(z_1)\hat{\mathcal{S}}_{-}(z_2)|\psi(t)\rangle$ and $E_bS=\langle 0|\hat{\mathcal{E}}_{b_+}(z_1)\hat{\mathcal{S}}_{-}(z_2)|\psi(t)\rangle$, which in the frequency domain, to the lowest order of frequency $\omega$, are governed by
\begin{align}
\label{eq:eq6}
-i\partial_{z_1}\begin{pmatrix}
E_aS\\E_bS
\end{pmatrix}
=\begin{pmatrix}
\mathcal{V}_0+\omega(1+\mathcal{V}_1)/v&g\sec\theta/c\\
g\sec\theta/v&\omega/v\end{pmatrix}\begin{pmatrix}
E_aS\\E_bS
\end{pmatrix},
\end{align}
with $\mathcal{V}_0=ig_p^2/\gamma c(1-i\Omega^2/\gamma V)$, $\mathcal{V}_1=g_p^2V[(1+\gamma^2/\Omega^2)V-2i\gamma]\cos^2\theta/(\Omega^2+i\gamma V)^2$, and $v=c\cos^2\theta$. In the absence of interaction ($V=0$), the coupling between $E_aS$ and $E_bS$ is momentum-matched. With the interaction $V\neq0$, the strong dissipative interaction $\mathcal{V}_0\approx ig_p^2/\gamma c$ inside the EIT blockade radius $z_b$ [$V(z_b)=\Omega^2/\gamma$] destroys this matching condition, and the conversion of $E_bS$ to $E_aS$ is blockaded if $g_p^2/\gamma c\gg g/v$. As a result of the blockade, the state evolution is restricted to the interaction-free subspace $E_bS$, endowed with a small decay coefficient $g^2\gamma c/g_p^2v^2$ and a slightly modified group velocity $v/[1+g^2(\Omega^2+\gamma^2)/(g_p^2\Omega^2\cos^2\theta)]$. Thus, the strong interaction-induced dissipations are suppressed.

For atomic coupling scheme, the temporal modulation $g=g(t)$ is chosen. In this case, we introduce two-photon wavefunctions $\Psi_\mu S$, $\Phi_\mu S$, $P_\mu S$ to describe DSP, BSP, and the excited state field of mode $\mu_+$ ($\mu=a,b$). Given a $V$, the dynamics can be described in the momentum-space (transforming from $z_1$ to $k$). Up to the first order of $k$, the equations of motion are approximately given by
\begin{align}
\label{eq:eq7}
i\partial_{t}\begin{pmatrix}
\hat{\psi}\\ \Psi_bS
\end{pmatrix}
=\begin{pmatrix}
\hat{\mathcal{V}}(k)+k\hat{v}&\hat{g}\\
\hat{g}^\dagger&kc\cos^2\theta\end{pmatrix}\begin{pmatrix}
\hat{\psi}\\ \Psi_bS
\end{pmatrix},
\end{align}
where $\hat{\psi}=(\Psi_aS,\Phi_aS, P_aS)^T$, $\hat{g} = (g, -g\cot\theta, 0)^T$, $\hat{v}=\mathbf{diag}(c\cos^2\theta, c\sin^2\theta, 0)$, and the effective potential $\hat{\mathcal{V}}(k)$ is a matrix of $k$ (see Appendix \ref{app:appB}). For the noninteracting case, Eq.~(\ref{eq:eq7}) describes an energy resonant coupling between $\Psi_bS$ and $\Psi_aS$. With a strong interaction $V$, the dynamics of $\hat{\psi}$ governed by $\hat{\mathcal{V}}(k)$ carry dissipative component. However, if the energy detuning induced by $\hat{\mathcal{V}}(k)$ ($\sim V\sin^2\theta$) is much stronger than the coupling induced by $\hat{g}$ ($\sim g$), the coupling between $\Psi_bS$ and the interacting space $\hat{\psi}$ can be successfully blockaded, with the range characterized by the coupling blockade radius $r_b$ defined as $V(r_b)\sin^2\theta=2g$ (to be distinguished from $z_b$). Inside $r_b$, the dynamics can be described in the $\Psi_bS$ space with hardly modified eigen energy and group velocity, where interaction-induced dissipations are largely suppressed.

For spatial and temporal modulation of the coupling considered above, blockade originates respectively from the interaction caused momentum-mismatch and energy off-resonance. We will focus on temporal modulation with atomic coupling scheme in this study, since it supports richer applications (see Appendix \ref{app:appC}). In this case, the coupling $g$ is turned on when two photons are close to each other, and is turned off before they completely separate. Thus, for photons with compressed pulse length $\sigma$, an ideal blockade requires $\sigma+\frac{1}{2}(v_++v_-)t<r_b$ (which reduces to $\sigma<r_b$ for copropagating photons), as otherwise photons separated by more than $r_b$ apart do not feel the blockade. Since EIT displays a finite transparency window, a smaller $\sigma$ will cause a proportionally larger linear loss. This leads to a tradeoff between ideal blockade and linear loss, which can be relaxed by increasing the blockade optical depth $d_b=g_p^2z_b/\gamma c$ of the system (see Appendix \ref{app:appD}).

Although we focus on Rydberg EIT system, the extension to other photonic systems with long-range interactions is straightforward, e.g., waveguide QED systems \cite{douglas2016photon,shahmoon2016highly}. In fact, the specific forms of interactions depend on implementations, but the generic form of Eq.~(\ref{eq:eq3}) applies, as long as the coupling between the interaction-free subspace and $\Psi_{a_+,a_-}$ is blockaded (because the interacting detail of $\Psi_{a_+,a_-}$ becomes unimportant in this case). Thus, we will use the generic model to illustrate the basic physics in the next section, while in numerical simulations we consider the full dynamics governed by complete Rydberg EIT coupled equations (see Appendix \ref{app:appB}).

The calculations are based on solving a set of coupled one-dimensional partial differential equations for two-particle wavefunctions. In the simulation, we assume a uniform atomic density and a square pulse of the coupling field $\Omega_c(t)$ for simplicity. We also neglect the spontaneous decay or dephasing of the Rydberg spin-wave field, since the propagation time inside the atomic ensemble is small enough (see Appendix \ref{app:appD}).

\section{Applications}\label{sec:sec4}
\subsection{Single-photon quantum switch}\label{sec:sec4A}

\begin{figure}
\centering
\includegraphics[width=\linewidth]{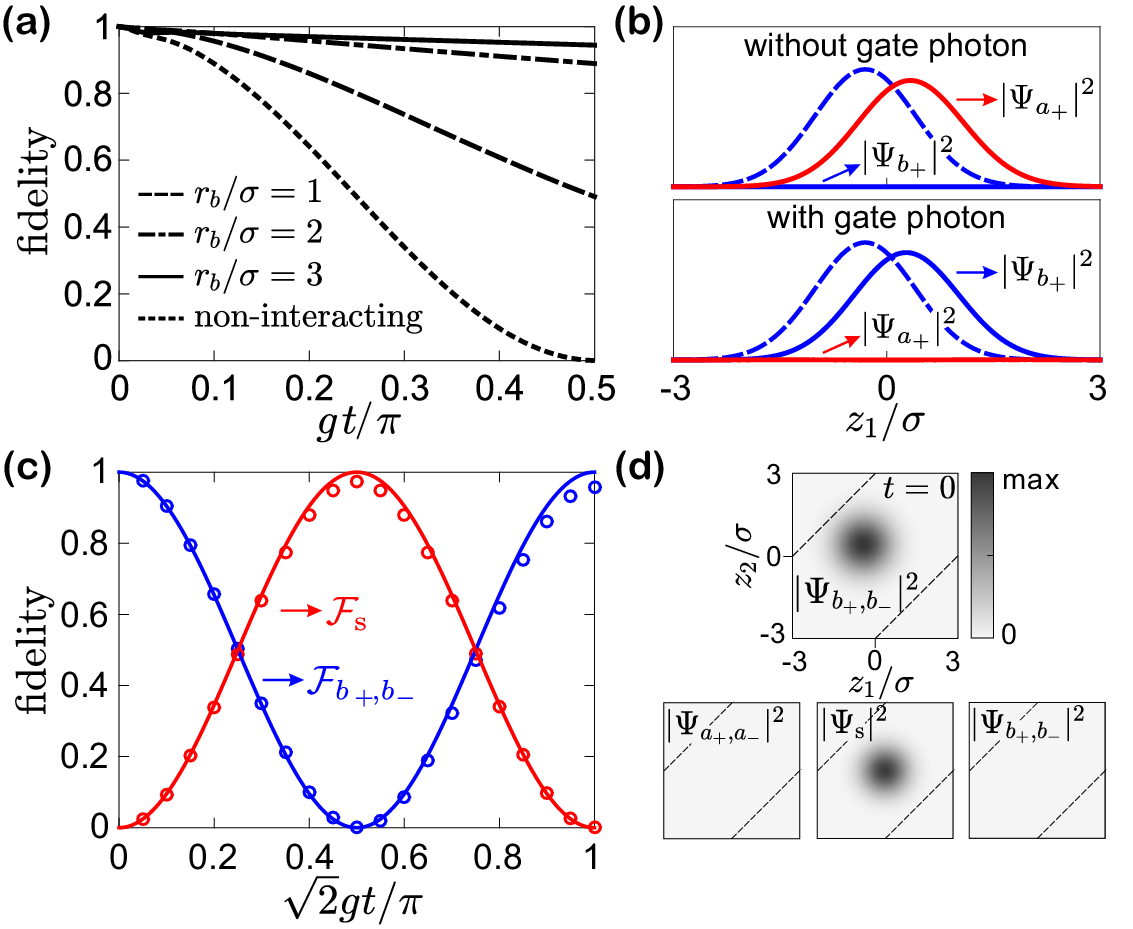}
\caption{(a) Evolution of the switching fidelity for different pulse lengths. (b) Evolution of the target photon wavefunction in mode $a_+$ ($|\Psi_{a_+}(z_1)|^2$) and $b_+$ ($|\Psi_{b_+}(z_1)|^2$) with $\sigma=r_b/3$ and $gt=\pi/2$, where the dashed line represents the initial wavefunction in mode $b_+$. (c) Evolution of fidelities $\mathcal{F}_{\mathrm{s}}$ and $\mathcal{F}_{b_+,b_-}$. The circles denote numerical calculations and the solid lines represent expected sinusoidal oscillation. (d) Two-photon wavefunctions of the entangled state produced ($\mathcal{F}_\mathrm{s}=97.26\%$) at $\sqrt{2}gt=\pi/2$ with $\sigma=r_b/3$. The simulation is carried out for $\Omega/2\pi=5$~MHz, $g_p/2\pi=20$~GHz, $\gamma/2\pi=3$~MHz, $z_b=13.8\mu\mathrm{m}$, and $r_b=16.5\mu\mathrm{m}$. The results are insensitive to the choices of parameters over a broad range. We consider a small group velocity mismatch in (a) and (b) by taking $\Omega_2=\Omega=0.9\Omega_1$, and neglect the linear loss of mode $b_\pm$ in (c) and (d) (see Appendix \ref{app:appB}).}
\label{fig:fig2}
\end{figure}

As in the previous section, if we take the backward photon to be a stored gate excitation, the equation of motion described by the generic model now reduces to
\begin{equation}
\label{eq:eq8}
i\partial_\xi\begin{pmatrix}
\Psi_{a_+a_-}\\\Psi_{b_+a_-}
\end{pmatrix}=\begin{pmatrix}
\mathcal{V}(r)&g\\
g&0
\end{pmatrix}\begin{pmatrix}
\Psi_{a_+a_-}\\\Psi_{b_+a_-}
\end{pmatrix}.
\end{equation}
In the absence of $\mathcal{V}$, the forward photon undergoes complete Rabi oscillation between mode $a_+$ and $b_+$. As explained previously, if $|\mathcal{V}(r)|\gg g$ is satisfied for all $r$ within the spread range of the photon, the coupling between $b_+$ and $a_+$ becomes blockaded everywhere. Such a phenomenon can be used to implement a single-photon optical switch. For a coupling duration $t=\pi/2g$, the target photon in mode $b_+$ is transformed to mode $a_+$ in the absence of the gate photon. This mode conversion becomes blockaded if the gate photon is present. As shown in Fig.~\ref{fig:fig2}(a), the switching fidelity becomes higher when the blockade radius $r_b$ is sufficiently longer compared with $\sigma$. Figure \ref{fig:fig2}(b) displays a typical calculated switching function for $r_b/\sigma=3$. At $t=\pi/g$, the system functions as a $\pi$-phase gate, which gives $\Psi_{b_+a_-}(z_1,z_2,t)=e^{i\pi}\Psi_{b_+a_-}(z_1-vt,z_2,0)$ for $\mathcal{V}=0$, and $\Psi_{b_+a_-}(z_1,z_2,t)=\Psi_{b_+a_-}(z_1-vt,z_2,0)$ for $|\mathcal{V}(\sigma)|\gg g$.

\subsection{Entangled state generation}\label{sec:sec4B}
To generate entanglement between two photonic modes, we take $a_-$ and $b_-$ to be backward propagating modes of $a_+$ and $b_+$ with $v_\pm=v$, $\pm\Delta k_\pm=\Delta k$, and $g_\pm=g$. Introducing the symmetric wavefunction $\tilde{\Psi}_\mathrm{s}=\left(\tilde{\Psi}_{a_+b_-}+\tilde{\Psi}_{b_+a_-}\right)/\sqrt{2}$, the Hamiltonian for the generic model reduces to
\begin{equation}
\label{eq:eq9}
\mathcal{H}=\begin{pmatrix}
2v\Delta k+\mathcal{V}(r)&\sqrt{2}g&0\\
\sqrt{2}g&v\Delta k&\sqrt{2}g\\
0&\sqrt{2}g&0
\end{pmatrix},
\end{equation}
and the state is given by $\psi=\left(\tilde{\Psi}_{a_+a_-},\tilde{\Psi}_\mathrm{s},\tilde{\Psi}_{b_+b_-}\right)^T$. As explained previously, strong interaction can block the excitation of $\Psi_{a_+,a_-}$, so that photons initially residing in state $|b_+,b_-\rangle$ can be efficiently transferred to an entangled state $|\Psi_\mathrm{s}\rangle=\left(|a_+,b_-\rangle + |b_+,a_-\rangle\right)/\sqrt{2}$. For the initial state $\tilde{\Psi}_{b_+b_-}(R,r,t=0)=\Psi_0(R,r)$, the state evolution is given by $\tilde{\Psi}_\mathrm{s}(R,r,t) = -i\Psi_0(R,r-2vt)\sin(\sqrt{2}gt)$, and $\tilde{\Psi}_{b_+b_-}(R,r,t) = \Psi_0(R,r-2vt)\cos(\sqrt{2}gt)$,
which corresponds to a Rabi oscillation between the initial and the entangled state at an enhanced rate $\sqrt{2}g$, in agreement with numerical results shown in Fig.~\ref{fig:fig2}(c). Thus, starting with a separable state $|b_+b_-\rangle$, an entangled photon pair appears with a high fidelity at $t=\pi/2\sqrt{2}g$ [see Fig.~\ref{fig:fig2}(d)].

\subsection{Nonlinear beam splitting}\label{sec:sec4C}
The input photonic state $(1/\sqrt{n!})\left[\int dz h(z)\hat{\Psi}_{b}^\dagger(z)\right]^{n}|0\rangle$ containing no photon in mode $a$ and $n$ copropagating photons in mode $b$ with identical wavefunction $h(z)$ is denoted by $|0,n\rangle$. The classical linear beam splitting coupling transforms $\hat{\Psi}_b(z)$ into $T\hat{\Psi}_b(z)+R\hat{\Psi}_a(z)$, whose output state $|\psi(t)\rangle=\sum_{m}\sqrt{n!/m!(n-m)!}T^{n-m}R^m|m,n-m\rangle$ inevitably includes multi-excitations of mode $a$.

\begin{figure}
\centering
\includegraphics[width=\linewidth]{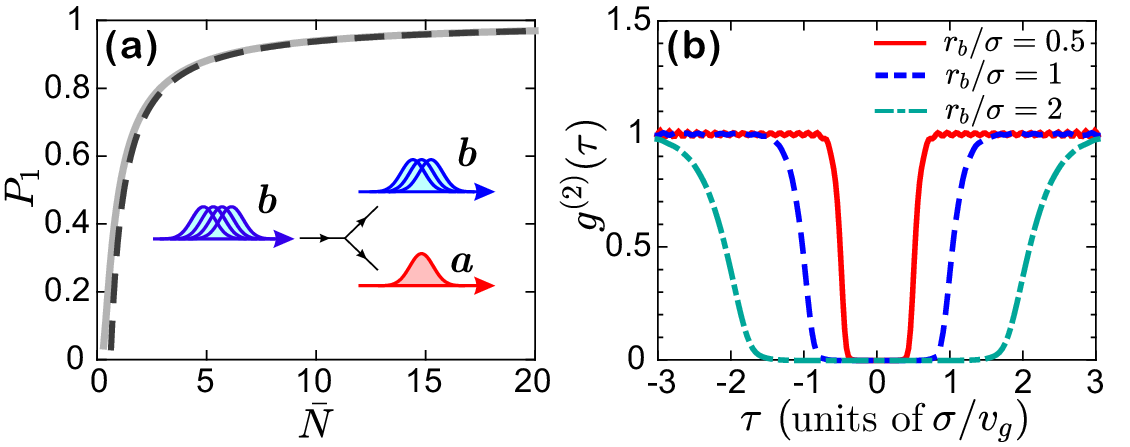}
\caption{(a) Single photon output efficiency as a function of the mean photon number $\bar{N}$ for a coherent state input. The solid and dashed lines represent the numerical summation results and the asymptotic solution $1-\pi^2/(16\bar{N})$, respectively. (b) Second order correlation functions of the output state for a weak coherent input at different values of pulse length $\sigma$. The parameters are the same as in Fig.~\ref{fig:fig2}(c).}
\label{fig:fig3}
\end{figure}

When a sufficiently strong $|\mathcal{V}(r)|\gg g\sqrt{n}$ is present for all $|r|\lesssim\sigma$, multi-excitations are blocked, and the initial state $|0,n\rangle$ is only effectively coupled to $|1,n-1\rangle$, giving rise to nonlinear beam splitting [see the inset of Fig.~\ref{fig:fig3}(a)]. Tracing out mode $b$, excitation probabilities for the vacuum $|0_a\rangle\langle0_a|$ and single-photon state $|1_a\rangle\langle1_a|$ are found to be $p_{0,n}=\cos^2(\sqrt{n}gt)$ and $p_{1,n}=\sin^2(\sqrt{n}gt)$, respectively. For an arbitrary incoming state with photon number distribution probability $f_n$, the single photon excitation probability becomes $P_1=\sum_nf_np_{1,n}$. For a coherent input of a mean photon number $\bar{N}$ ($f_n=\bar{N}^ne^{-\bar{N}}/n!$) and an operating time $t=\pi/(2\sqrt{\bar{N}}g)$, we find $P_1\approx1-\pi^2/(16\bar{N})$ when $\bar{N}$ is large [see Fig.~\ref{fig:fig3}(a)].

Such a nonlinear beam splitting can yield efficient single photon emission from a classical state. Different from the Rydberg EIT absorption protocol which always reduces the state purity \cite{gorshkov2013dissipative}, the single-photon quantum state obtained here is pure for ideal blockade. Its second order correlation function $g^{(2)}(\tau)$ for a weak coherent input state at $t=\pi/2g$ is shown in Fig.~\ref{fig:fig3}(b). As expected, the resulting anti-bunching region is distinguished by the blockade radius $r_b$, and a pulse with $\sigma<r_b$ can thus deterministically generate single-photons.

\subsection{Tunable dressed interaction}\label{sec:sec4D}
\begin{figure}
\centering
\includegraphics[width=\linewidth]{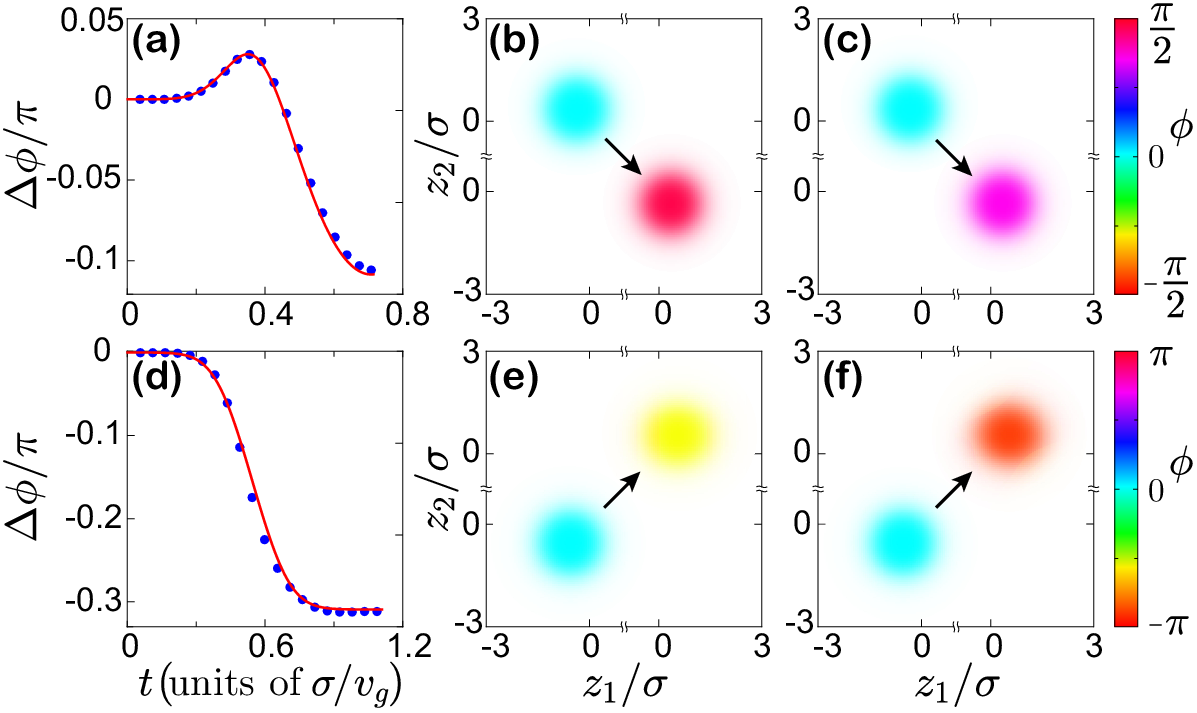}
\caption{Dressed interaction induced phase shifts $\Delta\phi=\phi_\mathrm{b}-\phi_\mathrm{n}$, with $\phi_\mathrm{b}$ and $\phi_\mathrm{n}$ the phase shift for the blockaded and non-interacting case, respectively. (a)-(c) show the counterpropagating case with non-adiabatic manipulations. $\Delta k=6.25/\sigma$ is used with all other parameters the same as in Fig.~\ref{fig:fig2}(c). At $t=0.71\sigma/v_g$, the fidelity reaches $92.69\%$ and $94.72\%$ for the non-interacting [Fig.~\ref{fig:fig4}(b)] and blockaded [Fig.~\ref{fig:fig4}(c)] case, respectively. (d)-(f) show the copropagating case with adibatic manipulations, with $\sigma=6.51\mu\mathrm{m}$, $\Delta k=13.20/\sigma$, and $g(t)/v_g\Delta k= [\tanh(5.78v_gt/\sigma-2.2)-\tanh(5.78v_gt/\sigma-4.1)]/4$. At $t=1.09\sigma/v_g$, the fidelity becomes $93.04\%$ and $91.45\%$ for the non-interacting [Fig.~\ref{fig:fig4}(e)] and blockaded [Fig.~\ref{fig:fig4}(f)] cases, respectively. In (a) and (d), the solid lines and circles respectively denote analytical and numerical results. In (b), (c), (e), and (f), the color and the opacity reflect the phase and the modulus square of the wavefunctions, respectively.}
\label{fig:fig4}
\end{figure}

Although bare interactions remain for photons populating the interacting two-photon mode, off-resonant couplings with the interacting nonlinear mode can further contribute dressed interactions to the noninteracting linear mode, in the same manner as the Rydberg-dressing \cite{johnson2010interactions,jau2015entangling}. One particular advantage is that in the blockade regime ($|\mathcal{V}(r<\sigma)|\gg g,v\Delta k$), these dressed interactions are uniform within the spread range of the photonic wavefunction. Its strength for no photon populating the interacting mode is
\begin{equation}
J_{bb}=-\frac{v\Delta k}{2}+\sqrt{(v\Delta k)^2+4g^2}-\frac{1}{2}\sqrt{(v\Delta k)^2+8g^2},\label{eq:eq10}
\end{equation}
and the strength for only one photon populating the interacting mode is
\begin{equation}
J_{ab}=-\frac{v\Delta k}{2}+\frac{1}{2}\sqrt{(v\Delta k)^2+8g^2}.\label{eq:eq11}
\end{equation}
In the blockade regime, the dressed interaction strength only depends on tunable parameters $\Delta k$ and $g$, which enables the construction of a tunable controlled-phase gate. Figure \ref{fig:fig4} presents such a plausible implementation with Rydberg EIT system, where in Figs.~\ref{fig:fig4}(a)-\ref{fig:fig4}(c), the interaction between two counterpropagating photons initially in the state $|b_+,b_-\rangle$ is considered. Turning on the coupling $g$ when the two photons enter the blockade region and turning it off before they leave, a phase accumulation $\Delta\phi$ can be obtained. For this nonadiabatic manipulation with moderate dressing ($g\lesssim v\Delta k$), a high-fidelity gate requires a suitably controlled operation time. An alternative choice comes from employing copropagating photons with adiabatic manipulations [Figs.~\ref{fig:fig4}(d)-\ref{fig:fig4}(f)], where the initial state $|a,b\rangle$ is kept in one of the dressed state when $g$ is adiabatically ramped up and down. The wavefunction $\Psi_{ab}$ picks up a dynamic phase $\Delta\phi=-\int dt^\prime J_{ab}(t^\prime)$ when $g$ is arranged to change along a closed path with no net flux. Comparisons between the non-interacting calculations [Figs.~\ref{fig:fig4}(b) and \ref{fig:fig4}(e)] and the blockaded results [Figs.~\ref{fig:fig4}(c) and \ref{fig:fig4}(f)] clearly demonstrate the dressed interaction induced phase shifts.

\section{Summary}\label{sec:sec5}
We show linearly coupled photonic systems with long-range interactions can exhibit coupling blockade for propagating photons. Several rudimentary quantum information protocols facilitated by such a blockade are discussed, including single-photon switching, entangled photon-pair generation, efficient single-photon emission, and photonic phase gate. Our proposal can be realized in photonic waveguide \cite{langbecker2017rydberg} or in atomic vapor \cite{Li2016quantum}, and the detailed experimental requirements taking into account the linear EIT loss, imperfect blockade, and group velocity mismatch are discussed in the Appendix \ref{app:appD}. In addition to photonic quantum state manipulation and their possible extensions to quantum information processing, the mechanism we discuss opens an exciting avenue towards constructing tunable and lossless photon-photon interactions in quantum nonlinear optics.

\begin{acknowledgments}
This work is supported by the National Key R$\&$D Program of China (Grant No.~2018YFA0306503 and No.~2018YFA0306504) and by NSFC (Grant No.~11574177, No.~91636213, No.~11654001, No.~11674390, and No.~91736106). Fan Yang acknowledges helpful discussions with Chuyang Shen, Cheng Chen, Dr.~Shen Dong, and Dr.~Lin Li.
\end{acknowledgments}

\begin{appendix}
\begin{widetext}
\section{Equations of motion for counterpropagating and copropagating photons}\label{app:appA}
In this section, we provide a detailed derivation of the Schrodinger-like equations for describing the evolution of the two-photon wavefunctions in the generic model.

In the case of two counterpropagating photons, the model Hamiltonian is described by Eqs.~(\ref{eq:eq1})-(\ref{eq:eq3}). With the commutation relation $[\Psi_\mu(z,t),\Psi_\nu^\dagger(z^\prime,t)]=\delta_{\mu\nu}\delta(z-z^\prime)$ for bosonic particles, the Heisenberg equations for the field operators are given by
\begin{align}
\partial_t\hat{\Psi}_{a_+}(z) &=-v_{+}\partial_z\hat{\Psi}_{a_+}(z)-ig_+\hat{\Psi}_{b_+}(z)e^{i\Delta k_+z}-i\int dz^\prime\mathcal{V}(z-z^\prime)\hat{\Psi}_{a_-}^\dagger(z^\prime)\hat{\Psi}_{a_-}(z^\prime)\hat{\Psi}_{a_+}(z),\label{eq:eqA1}\\
\partial_t\hat{\Psi}_{b_+}(z) &=-v_{+}\partial_z\hat{\Psi}_{b_+}(z)-ig_+\hat{\Psi}_{a_+}(z)e^{-i\Delta k_+z},\label{eq:eqA2}\\
\partial_t\hat{\Psi}_{a_-}(z) &= v_{-}\partial_z\hat{\Psi}_{a_-}(z)-ig_-\hat{\Psi}_{b_-}(z)e^{i\Delta k_-z}-i\int dz^\prime\mathcal{V}(z^\prime-z)\hat{\Psi}_{a_+}^\dagger(z^\prime)\hat{\Psi}_{a_+}(z^\prime)\hat{\Psi}_{a_-}(z),\label{eq:eqA3}\\
\partial_t\hat{\Psi}_{b_-}(z) &= v_{-}\partial_z\hat{\Psi}_{b_-}(z)-ig_-\hat{\Psi}_{a_-}(z)e^{-i\Delta k_-z},\label{eq:eqA4}
\end{align}
In the Schrodinger picture, the state of the system can be written as
\begin{align}
|\psi(t)\rangle =& \iint dz_1dz_2\Psi_{a_+a_-}(z_1,z_2,t)\hat{\Psi}_{a_+}^\dagger(z_1)\hat{\Psi}_{a_-}^\dagger(z_2)|0\rangle + \iint dz_1dz_2\Psi_{a_+b_-}(z_1,z_2,t)\hat{\Psi}_{a_+}^\dagger(z_1)\hat{\Psi}_{b_-}^\dagger(z_2)|0\rangle\nonumber\\
&+\iint dz_1dz_2\Psi_{b_+a_-}(z_1,z_2,t)\hat{\Psi}_{b_+}^\dagger(z_1)\hat{\Psi}_{a_-}^\dagger(z_2)|0\rangle + \iint dz_1dz_2\Psi_{b_+b_-}(z_1,z_2,t)\hat{\Psi}_{b_+}^\dagger(z_1)\hat{\Psi}_{b_-}^\dagger(z_2)|0\rangle,\label{eq:eqA5}
\end{align}
where $\Psi_{\mu_+\nu_-}(z_1,z_2,t)=\langle0|\hat{\Psi}_{\mu_+}(z_1)\hat{\Psi}_{\nu_-}(z_2)|\psi(t)\rangle$ denotes a two-photon wavefunction. With the Heisenberg equations (\ref{eq:eqA1})-(\ref{eq:eqA4}), the equation of motion for the two-photon wavefunctions becomes
\begin{align}
\partial_t\Psi_{a_+a_-} &=-v_{+}\partial_{z_1}\Psi_{a_+a_-}-ig_+\Psi_{b_+a_-}e^{i\Delta k_+z_1}+v_{-}\partial_{z_2}\Psi_{a_+a_-}-ig_-\Psi_{a_+b_-}e^{i\Delta k_-z_2}-i\mathcal{V}(z_1-z_2)\Psi_{a_+a_-},\label{eq:eqA6}\\
\partial_t\Psi_{a_+b_-} &=-v_{+}\partial_{z_1}\Psi_{a_+b_-}-ig_+\Psi_{b_+b_-}e^{i\Delta k_+z_1}+v_{-}\partial_{z_2}\Psi_{a_+b_-}-ig_-\Psi_{a_+a_-}e^{-i\Delta k_-z_2},\label{eq:eqA7}\\
\partial_t\Psi_{b_+a_-} &=-v_{+}\partial_{z_1}\Psi_{b_+a_-}-ig_+\Psi_{a_+a_-}e^{-i\Delta k_+z_1}+v_{-}\partial_{z_2}\Psi_{b_+a_-}-ig_-\Psi_{b_+b_-}e^{i\Delta k_-z_2},\label{eq:eqA8}\\
\partial_t\Psi_{b_+b_-} &=-v_{+}\partial_{z_1}\Psi_{b_+b_-}-ig_+\Psi_{a_+b_-}e^{-i\Delta k_+z_1}+v_{-}\partial_{z_2}\Psi_{b_+b_-}-ig_-\Psi_{b_+a_-}e^{-i\Delta k_-z_2}.\label{eq:eqA9}
\end{align}
Transforming the wavefunctions into the rotating frame, according to
\begin{equation}
\Psi_{a_+,a_-}=\tilde{\Psi}_{a_+,a_-}e^{i(\Delta k_+z_1+\Delta k_-z_2)}, \Psi_{a_+,b_-}=\tilde{\Psi}_{a_+,b_-}e^{i\Delta k_+z_1}, \Psi_{b_+,a_-}=\tilde{\Psi}_{b_+,a_-}e^{i\Delta k_-z_2}, \Psi_{b_+,b_-}=\tilde{\Psi}_{b_+,b_-},\label{eq:eqA10}
\end{equation}
Eqs.~(\ref{eq:eqA6})-(\ref{eq:eqA9}) are simplified to
\begin{equation}
(\partial_t+v_{+}\partial_{z_1}-v_{-}\partial_{z_2})\begin{pmatrix}
\tilde{\Psi}_{a_+a_-}\\\tilde{\Psi}_{a_+b_-}\\\tilde{\Psi}_{b_+a_-}\\\tilde{\Psi}_{b_+b_-}
\end{pmatrix}=-i
\begin{pmatrix}
v_{+}\Delta k_+-v_{-}\Delta k_-+\mathcal{V}(z_1-z_2)&g_-&g_+&0\\
g_-&v_{+}\Delta k_+&0&g_+\\
g_+&0&-v_{-}\Delta k_-&g_-\\
0&g_+&g_-&0
\end{pmatrix}\begin{pmatrix}
\tilde{\Psi}_{a_+a_-}\\\tilde{\Psi}_{a_+b_-}\\\tilde{\Psi}_{b_+a_-}\\\tilde{\Psi}_{b_+b_-}
\end{pmatrix}.\label{eq:eqA11}
\end{equation}
Further introducing center-of-mass and relative coordinates $R=(v_{-}z_1+v_{+}z_2)/(v_{+}+v_{-})$ and $\rho=z_1-(v_{-}/v_{+})z_2$, and moving into the retarded time frame with $\tau=t-\rho/v_\mathrm{eff}$ and $\xi=t$, we can obtain Eq.~(\ref{eq:eq5}) in the main text.

For the case of two copropagating photons, the Hamiltonian of the system is $\hat{H}=\hat{H}_0+\hat{H}_\mathrm{c}+\hat{H}_\mathrm{int}$, where
\begin{align}
\hat{H}_0 =& -iv\int dz\hat{\Psi}_{a}^\dagger(z) \partial_z\hat{\Psi}_{a}(z)-iv\int dz\hat{\Psi}_{b}^\dagger(z) \partial_z\hat{\Psi}_{b}(z), \label{eq:eqA12}\\
\hat{H}_\mathrm{c} =& \int dz g\hat{\Psi}_{a}^\dagger(z)\hat{\Psi}_{b}(z)e^{i\Delta k z}+\mathrm{H.c.},\label{eq:eqA13}\\
\hat{H}_\mathrm{int} =&\frac{1}{2}\int dz dz^\prime \mathcal{V}(z-z^\prime)\hat{\Psi}_{a}^\dagger(z)\hat{\Psi}_{a}^\dagger(z^\prime)\hat{\Psi}_{a}(z^\prime)\hat{\Psi}_{a}(z).\label{eq:eqA14}
\end{align}
The state of the system is described by the two-photon wavefunction $\Psi_{\mu\nu}(z_1,z_2,t)=\langle0|\hat{\Psi}_{\mu}(z_1)\hat{\Psi}_{\nu}(z_2)|\psi(t)\rangle$, i.e.,
\begin{align}
|\psi(t)\rangle =& \frac{1}{2}\iint dz_1dz_2\Psi_{aa}(z_1,z_2,t)\hat{\Psi}_{a}^\dagger(z_1)\hat{\Psi}_{a}^\dagger(z_2)|0\rangle + \frac{1}{2}\iint dz_1dz_2\Psi_{bb}(z_1,z_2,t)\hat{\Psi}_{b}^\dagger(z_1)\hat{\Psi}_{b}^\dagger(z_2)|0\rangle\nonumber\\
&+  \iint dz_1dz_2\Psi_{ab}(z_1,z_2,t)\hat{\Psi}_{a}^\dagger(z_1)\hat{\Psi}_{b}^\dagger(z_2)|0\rangle.\label{eq:eqA15}
\end{align}
Employing the same procedure as developed for the counterpropagating case, the equations of motion in the moving and rotating frame can be simplified to
\begin{equation}
(\partial_t+v\partial_{R})\begin{pmatrix}
\tilde{\Psi}_{aa}\\\tilde{\Psi}_{ab}\\\tilde{\Psi}_{ba}\\\tilde{\Psi}_{bb}
\end{pmatrix}=-i
\begin{pmatrix}
2v\Delta k+\mathcal{V}(r)&g&g&0\\
g&v\Delta k&0&g\\
g&0&v\Delta k&g\\
0&g&g&0
\end{pmatrix}\begin{pmatrix}
\tilde{\Psi}_{aa}\\\tilde{\Psi}_{ab}\\\tilde{\Psi}_{ba}\\\tilde{\Psi}_{bb}
\end{pmatrix}.\label{eq:eqA16}
\end{equation}

In the calculations, the fidelity $\mathcal{F}_{\mu\nu}$ and the phase $\phi_{\mu\nu}$ to the state $\hat{U}_0(t,0)\iint dz_1dz_2 \Psi_0(z_1,z_2)\hat{\Psi}_{\mu_+}^\dagger(z_1)\hat{\Psi}_{\nu_-}^\dagger(z_2)|0\rangle$ is determined by
\begin{equation}
\sqrt{\mathcal{F}_{\mu\nu}}e^{i\phi_{\mu\nu}}=\int dz_1 dz_2\Psi_0^*(z_1-v_+t,z_2+v_-t)\Psi_{\mu_+\nu_-}(z_1,z_2,t),\label{eq:eqA17}
\end{equation}
for two conterpropagating photons (and analogously for two copropagating photons), where $\hat{U}_0(t,0)$ denotes the noninteracting unitary time-evolution operator, and $\Psi_0(z_1,z_2)$ represents the normalized initial wavefunction (at $t=0$). The fidelity $\mathcal{F}_\mathrm{s}$ for the symmetric entangled state is obtained by replacing the wavefunction $\Psi_{\mu_+\nu_-}(z_1,z_2,t)$ with $\frac{1}{\sqrt{2}}\left[\Psi_{a_+b_-}(z_1,z_2,t)+\Psi_{b_+a_-}(z_1,z_2,t)\right]$ in Eq.~(\ref{eq:eqA17}).

\section{Rydberg EIT implementation}\label{app:appB}
In this section, we describe how to map the Rydberg EIT system to the generic model derivated in the previous section. In a Rydberg EIT system \cite{gorshkov2011photon}, the evolution of light and atomic excitations can be described by bosonic operators $\hat{\mathcal{E}}_{a_\pm}^\dagger(z,t)$, $\hat{\mathcal{P}}_{a_\pm}^\dagger(z)$, and $\hat{\mathcal{S}}_{a_\pm}^\dagger(z)$, which create a photon, an atomic spin-wave in intermediate state $|e\rangle$, and an atomic spin-wave in Rydberg state $|r\rangle$, respectively. We take the dark state polariton to be the interacting mode $a_\pm$, i.e., $\hat{\Psi}_{a_\pm}(z)=\cos\theta\hat{\mathcal{E}}_{a_\pm}(z)- \sin\theta\hat{\mathcal{S}}_{a_\pm}(z)$ with $\tan\theta=g_p/\Omega$ (see the main text).

To gain the essential physics of the coupling blockade, we first consider the simple case with the target photon stored in an auxiliary Rydberg state, i.e., we take $\hat{\Psi}_{a_-}(z)=\hat{\mathcal{S}}_-(z)$. For the photonic coupling scheme, the coupling between $a_+$ and another linear mode $b_+$ is obtained through photonic coupling term $g\sec\theta\hat{\mathcal{E}}_{a_+}^\dagger(z)\hat{\mathcal{E}}_{b_+}(z)+\mathrm{H.c.}$, and the equations of motion for the field operators are given by
\begin{align}
\partial_t\hat{\mathcal{E}}_{a_+}(z)&=-c\partial_z\hat{\mathcal{E}}_{a_+}(z)+ig_p\hat{\mathcal{P}}_{a_+}(z)-ig\sec\theta\hat{\mathcal{E}}_{b_+}(z),\label{eq:eqB1}\\
\partial_t\hat{\mathcal{P}}_{a_+}(z)&=-\gamma\hat{\mathcal{P}}_{a_+}(z)+ig_p\hat{\mathcal{E}}_{a_+}(z)+i\Omega\hat{\mathcal{S}}_{a_+}(z),\label{eq:eqB2}\\
\partial_t\hat{\mathcal{S}}_{a_+}(z)&=i\Omega\hat{\mathcal{P}}_{a_+}(z)-i\int dz^\prime V(z-z^\prime)\hat{\mathcal{S}}_-^\dagger(z^\prime)\hat{\mathcal{S}}_-(z^\prime)\hat{\mathcal{S}}_{a_+}(z),\label{eq:eqB3}\\
\partial_t\hat{\mathcal{E}}_{b_+}(z)&=-v\partial_z\hat{\mathcal{E}}_{b_+}(z)-ig\sec\theta\hat{\mathcal{E}}_{a_+}(z),\label{eq:eqB4}\\
\partial_t\hat{\mathcal{S}}_-(z)&=-i\int dz^\prime V(z-z^\prime)\hat{\mathcal{S}}_{a_+}^\dagger(z^\prime)\hat{\mathcal{S}}_{a_+}(z^\prime)\hat{\mathcal{S}}_-(z).\label{eq:eqB5}
\end{align}
In Eq.~(\ref{eq:eqB2}), we neglect the Langevin noise terms associated with the decay of the field operator $\hat{\mathcal{P}}_{a_+}(z)$, since they do not affect the calculation of the two-photon wavefunction \cite{gorshkov2011photon}. Introducing the two-particle wavefunction as in Appendix \ref{app:appA}, the state $|\psi(t)\rangle$ of the system can be described by $\psi(z_1,z_2,t)=\left(E_aS,P_aS,S_aS,E_bS\right)^{T}$ with each element being a two-particle wavefunction component (e.g., $E_aS=\langle0|\hat{\mathcal{E}}_{a_+}(z_1)\hat{\mathcal{S}}_{-}(z_2)|\psi(t)\rangle$), whose dynamics are governed by
\begin{align}
\partial_tE_aS(z_1,z_2)&=-c\partial_{z_1}E_aS(z_1,z_2)+ig_pP_aS(z_1,z_2)-ig\sec\theta E_bS(z_1,z_2),\label{eq:eqB6}\\
\partial_tP_aS(z_1,z_2)&=-\gamma P_aS(z_1,z_2)+ig_pE_aS(z_1,z_2)+i\Omega S_aS(z_1,z_2),\label{eq:eqB7}\\
\partial_tS_aS(z_1,z_2)&=i\Omega P_aS(z_1,z_2)-i V(z_2-z_1)S_aS(z_1,z_2),\label{eq:eqB8}\\
\partial_tE_bS(z_1,z_2)&=-v\partial_{z_1}E_bS(z_1,z_2)-ig\sec\theta E_aS(z_1,z_2)\label{eq:eqB9}.
\end{align}
In the photonic coupling scheme, the passive modulation of the coupling $g$ is straighforwardly implemented, i.e., $g = g(z)$. Thus, we solve Eqs.~(\ref{eq:eqB6})-(\ref{eq:eqB9}) in the frequency domain by applying the Fourier transform of time $t$, i.e., introducing $\psi(z_1,z_2,\omega)=\int dt e^{i\omega t}\psi(z_1,z_2,t)$. Up to the first order of $\omega$, the equations of motion simplify to Eq.~(\ref{eq:eq6}). In the blockade region with $|\mathcal{V}_0|\gg g/v$, $E_bS$ is only perturbatively coupled with $E_aS$, and the effective dynamics of $E_bS$ is governed by
\begin{equation}
\partial_{z_1}E_bS = -i\frac{g^2}{(v^2\mathcal{V}_0)}E_bS - \frac{v}{1+g^2\mathcal{V}_1/v^2\mathcal{V}_0^2}\partial_tE_bS.\label{eq:eqB10}
\end{equation}
Since $\mathcal{V}_0(r)\approx ig_p^2/\gamma c$ for $r<z_b$ and $\mathcal{V}_0(r)\approx 0$ for $r>z_b$, [with $z_b$ defined by $V(z_b)=\Omega^2/\gamma$], coupling blockade in this case requires $g_p^2/\gamma c\gg g/v$, and the blockade range is characterized by the EIT blockade radius $z_b$.

For the atomic coupling scheme, the dynamics of the system shown in Fig.~\ref{fig:fig1}(d) in the main text can be described by the field operators $\hat{\mathcal{E}}_{a_+}^\dagger(z)$, $\hat{\mathcal{P}}_{a_+}^\dagger(z)$, $\hat{\mathcal{S}}_{a_+}^\dagger(z)$, $\hat{\mathcal{E}}_{b_+}^\dagger(z)$, $\hat{\mathcal{P}}_{b_+}^\dagger(z)$, and $\hat{\mathcal{S}}_{b_+}^\dagger(z)$, which create a photon with polarization $a$, an atomic spin-wave in intermediate state $|e_-\rangle$, an atomic spin-wave in Rydberg state $|r\rangle$, a photon with polarization $b$, an atomic spin-wave in intermediate state $|e_+\rangle$, and an atomic spin-wave in ground state $|d\rangle$. The coupling between the nonlinear DSP $\hat{\Psi}_{a_+}(z)=\cos\theta_1\hat{\mathcal{E}}_{a_+}(z)- \sin\theta_1\hat{\mathcal{S}}_{a_+}(z)$ and the linear DSP $\hat{\Psi}_{b_+}(z)=\cos\theta_2\hat{\mathcal{E}}_{b_+}(z)- \sin\theta_2\hat{\mathcal{S}}_{b_+}(z)$ is obtained through atomic coupling term $g\csc\theta_1\csc\theta_2\hat{\mathcal{S}}_{a_+}^\dagger(z)\hat{\mathcal{S}}_{b_+}(z)+\mathrm{H.c.}$ with $\tan\theta_1=g_p/\Omega_1$ and $\tan\theta_2=g_p/\Omega_2$ (assuming $|e_+\rangle$ and $|e_-\rangle$ are of the same hyperfine manifold). The dynamics of these operators are governed by
\begin{align}
\partial_t\hat{\mathcal{E}}_{a_+}(z)&=-c\partial_z\hat{\mathcal{E}}_{a_+}(z)+ig_p\hat{\mathcal{P}}_{a_+}(z),\label{eq:eqB11}\\
\partial_t\hat{\mathcal{P}}_{a_+}(z)&=-\gamma\hat{\mathcal{P}}_{a_+}(z)+ig_p\hat{\mathcal{E}}_{a_+}(z)+i\Omega_1\hat{\mathcal{S}}_{a_+}(z),\label{eq:eqB12}\\
\partial_t\hat{\mathcal{S}}_{a_+}(z)&=i\Omega_1\hat{\mathcal{P}}_{a_+}(z)-ig\csc\theta_1\csc\theta_2\hat{\mathcal{S}}_{b_+}(z,t)-i\int dz^\prime V(z-z^\prime)\hat{\mathcal{S}}_-^\dagger(z^\prime)\hat{\mathcal{S}}_-(z^\prime)\hat{\mathcal{S}}_{a_+}(z),\label{eq:eqB13}\\
\partial_t\hat{\mathcal{E}}_{b_+}(z)&=-c\partial_z\hat{\mathcal{E}}_{b_+}(z)+ig_p\hat{\mathcal{P}}_{b_+}(z),\label{eq:eqB14}\\
\partial_t\hat{\mathcal{P}}_{b_+}(z)&=-\gamma\hat{\mathcal{P}}_{b_+}(z)+ig_p\hat{\mathcal{E}}_{b_+}(z)+i\Omega_2\hat{\mathcal{S}}_{b_+}(z),\label{eq:eqB15}\\
\partial_t\hat{\mathcal{S}}_{b_+}(z)&=i\Omega_2\hat{\mathcal{P}}_{b_+}-ig\csc\theta_1\csc\theta_2\hat{\mathcal{S}}_{a_+}(z,t),\label{eq:eqB16}\\
\partial_t\hat{\mathcal{S}}_-(z)&=-i\int dz^\prime V(z-z^\prime)\hat{\mathcal{S}}_{a_+}^\dagger(z^\prime)\hat{\mathcal{S}}_{a_+}(z^\prime)\hat{\mathcal{S}}_-(z),\label{eq:eqB17}
\end{align}
and the corresponding equations of motion for the two-particle wavefunction are given by
\begin{align}
\partial_tE_aS(z_1,z_2)&=-c\partial_{z_1}E_aS(z_1,z_2)+ig_pP_aS(z_1,z_2),\label{eq:eqB18}\\
\partial_tP_aS(z_1,z_2)&=-\gamma P_aS(z_1,z_2)+ig_pE_aS(z_1,z_2)+i\Omega_1 S_aS(z_1,z_2),\label{eq:eqB19}\\
\partial_tS_aS(z_1,z_2)&=i\Omega_1 P_aS(z_1,z_2)-ig\csc\theta_1\csc\theta_2S_bS(z_1,z_2) -iV(z_2-z_1)S_aS(z_1,z_2),\label{eq:eqB20}\\
\partial_tE_bS(z_1,z_2)&=-c\partial_{z_1}E_bS(z_1,z_2)+ig_pP_bS(z_1,z_2),\label{eq:eqB21}\\
\partial_tP_bS(z_1,z_2)&=-\gamma P_bS(z_1,z_2)+ig_pE_bS(z_1,z_2)+i\Omega_2 S_bS(z_1,z_2),\label{eq:eqB22}\\
\partial_tS_bS(z_1,z_2)&=i\Omega_2 P_bS(z_1,z_2)-ig\csc\theta_1\csc\theta_2S_aS(z_1,z_2).\label{eq:eqB23}
\end{align}
In the atomic coupling scheme, the active modulation of the coupling $g$ is available, i.e., $g=g(t)$. Thus, we can solve Eqs.~(\ref{eq:eqB18})-(\ref{eq:eqB23}) in the momentum space by applying the Fourier transform $\psi(k,z_2,t)=\int dz_1 e^{-ikz_1}\psi(z_1,z_2,t)$. Following Ref.~\cite{otterbach2013wigner}, we set $V$ as a constant to see how the blockade works in this case, while in the numerical simulation $V$ is taken to be position-dependent rigorously. Introducing the bright state polaritons (BSPs) $\hat{\Phi}_{a_+}(z)=\sin\theta_1\hat{\mathcal{E}}_{a_+}(z)+\cos\theta_1\hat{\mathcal{S}}_{a_+}(z)$ and $\hat{\Phi}_{b_+}(z)=\sin\theta_2\hat{\mathcal{E}}_{b_+}(z)+\cos\theta_2\hat{\mathcal{S}}_{b_+}(z)$, the evolution of the two-photon wavefunction $\psi(z_1,z_2,t)=\left(\Psi_aS,\Phi_aS,P_aS,\Psi_bS,\Phi_bS,P_bS\right)^T$ in $k$-space is governed by $i\partial_t\psi=\mathcal{H}\psi$, with $\mathcal{H}$ given by
\begin{equation}
\mathcal{H}=\begin{pmatrix}
ck\cos^2\theta_1+V\sin^2\theta_1 & \sin\theta_1\cos\theta_1(ck-V) & 0 & g & -g\cot\theta_2 & 0\\
\sin\theta_1\cos\theta_1(ck-V) & ck\sin^2\theta_1+V\cos^2\theta_1 & -\sqrt{g_p^2+\Omega_1^2} & -g\cot\theta_1 & g\cot\theta_1\cot\theta_2 & 0\\
0 & -\sqrt{g_p^2+\Omega_1^2} & -i\gamma & 0 & 0 & 0\\
g & -g\cot\theta_1 & 0 & ck\cos^2\theta_2 & ck\sin\theta_2\cos\theta_2 & 0\\
-g\cot\theta_2 & g\cot\theta_1\cot\theta_2 & 0 & ck\sin\theta_2\cos\theta_2 & ck\sin^2\theta_2 & -\sqrt{g_p^2+\Omega_2^2}\\
0 & 0 & 0 & 0 & -\sqrt{g_p^2+\Omega_2^2} & -i\gamma
\end{pmatrix}.\label{eq:eqB24}
\end{equation}
In the above, the small coupling $-g\cot\theta_2$ between $\Psi_aS$ and $(\Phi_bS,P_bS)$ can be neglected in the slow-light regime ($\cos\theta_{1/2}\ll1$), and $\Psi_bS$ decouples with $(\Phi_bS,P_bS)$ up to the first order of $k$. Thus, $(\Phi_bS,P_bS)$ can be dropped out of the dynamics if we neglect the linear loss of $\hat{\Psi}_{b+}(z)$, and the corresponding equations of motion reduce to
\begin{align}
i\partial_{t}\begin{pmatrix}
\hat{\psi}\\ \Psi_bS
\end{pmatrix}
=\begin{pmatrix}
\hat{\mathcal{V}}(k)+k\hat{v}&\hat{g}\\
\hat{g}^\dagger&kc\cos^2\theta_2\end{pmatrix}\begin{pmatrix}
\hat{\psi}\\ \Psi_bS
\end{pmatrix},\label{eq:eqB25}
\end{align}
where $\hat{\psi}=(\Psi_aS,\Phi_aS, P_aS)^T$, $\hat{g} = (g, -g\cot\theta_1, 0)^T$, $\hat{v}=\mathbf{diag}(c\cos^2\theta_1, c\sin^2\theta_1, 0)$, and the effective potential $\hat{\mathcal{V}}(k)$ is
\begin{equation}
\hat{\mathcal{V}}(k)=\begin{pmatrix}
ck\cos^2\theta_1+V\sin^2\theta_1 & \sin\theta_1\cos\theta_1(ck-V) & 0\\
\sin\theta_1\cos\theta_1(ck-V) & ck\sin^2\theta_1+V\cos^2\theta_1 & -\sqrt{g_p^2+\Omega_1^2}\\
0 & -\sqrt{g_p^2+\Omega_1^2} & -i\gamma
\end{pmatrix}.\label{eq:eqB26}
\end{equation}
\end{widetext}
For the noninteracting case ($V=0$), $\Psi_a$ decouples with $(\Phi_aS,P_aS)$ up to the first order of $k$, and the coupling between $\Psi_bS$ and $(\Phi_aS,P_aS)$ is negligible in the slow-light regime, so that Eq.~(\ref{eq:eqB25}) can be simplified to
\begin{equation}
i\partial_{t}\begin{pmatrix}
\Psi_aS\\ \Psi_bS
\end{pmatrix}
=\begin{pmatrix}
kc\cos^2\theta_1&g\\
g&kc\cos^2\theta_2\end{pmatrix}\begin{pmatrix}
\Psi_aS\\ \Psi_bS
\end{pmatrix}.\label{eq:eqB27}
\end{equation}
For the interacting case ($V\neq0$), $\hat{\mathcal{V}}(k)$ is not diagonal, and the DSP field $\Psi_aS$ is strongly coupled with $\Phi_aS$ and $P_aS$ at $V\approx-g_p^2/(ck)$, which induces strong dissipative dyanmics of $\Psi_aS$. In general, the coupling between $\Psi_bS$ and $\hat{\psi}$ can be characterized by the modified eigen state $\Psi_bS^\prime$ through exact diagonalization (ED) of the Hamiltonian in Eq.~(\ref{eq:eqB25}), which is a linear superposition of $\Psi_bS$ and $\hat{\psi}$, i.e., $\Psi_bS^\prime=\sin\beta\Psi_bS+\cos\beta\hat{w}^\dagger\hat{\psi}$, with $\hat{w}$ being a weight matrix. The ideal blockade requires $\sin\beta\approx1$, and a smaller $\sin\beta$ indicates a less effective blockade. In the slow-light regime, the eigen energy of $\Psi_aS$ is characterized by $\mathcal{V}=V\sin^2\theta_1$, and thus the weight can be approximately expressed as
\begin{equation}
\label{eq:eqB28}
\sin^2\beta\approx 1-\frac{2(g/\mathcal{V})^2}{1+(2g/\mathcal{V})^2+\sqrt{1+(2g/\mathcal{V})^2}},
\end{equation}
which is verified by rigorous results from ED [see Fig.~\ref{fig:fig5}(a)]. With Eq.~(\ref{eq:eqB28}), we can introduce the coupling blockade radius $r_b$ defined as $\mathcal{V}(r_b)=2g$, at which we have $\sin\beta\approx0.9$. Such a blockade radius characterizes the effective blockade range, i.e., the coupling of $\Psi_bS$ to $\hat{\psi}$ can be effectively blockaded for $|r|<r_b$ region. Different from spatial modulation, the perturbative coupling with the interacting space induces negligible modifications of $\Psi_bS$ inside the blockade radius $r_b$. This is verified by the dispersion relation of the modified eigen state $\Psi_bS^\prime$ shown in Figs.~\ref{fig:fig5}(b) and \ref{fig:fig5}(c), as the group velocity is rarely changed and the dissipation is negligible during the operation time ($\sim g^{-1}$).

\begin{figure}
\centering
\includegraphics[width=\linewidth]{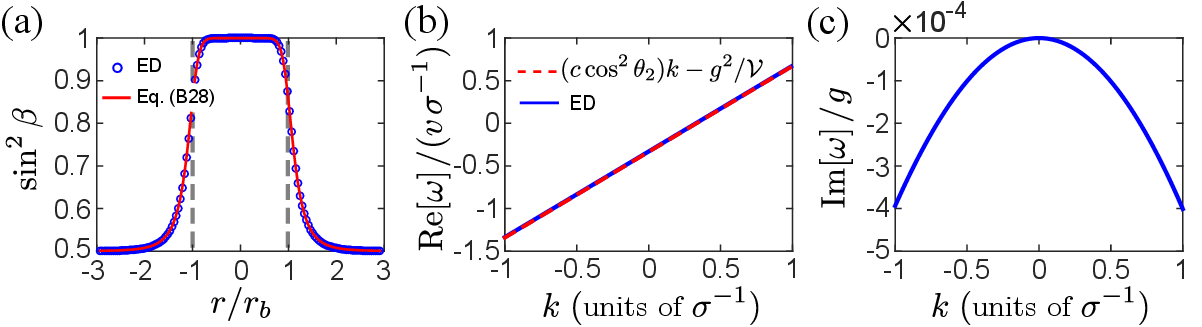}
\caption{(a) Weight of the $\Psi_bS$ component in the modified eigen state $\Psi_bS^\prime$ with $k=0.5/\sigma$. (b) and (c) show the dispersion relation of the modified eigen state $\Psi_bS^\prime$, in which (b) displays the real part of the eigen frequency $\omega$ and (c) plots the imaginary part of $\omega$ (at $r=0.8r_b$). The parameters are the same as in Fig.~\ref{fig:fig2} in the main text.}
\label{fig:fig5}
\end{figure}

Next, we consider the interaction between two propagating photons. We take counterpropagating photons with atomic coupling scheme as an example, while the generalization to the copropagating case and photonic coupling scheme is straightforward. For simplicity, we drop the field operators $\hat{\Phi}_{b_\pm}(z)$ and $\hat{\mathcal{P}}_{b_\pm}(z)$ out of the dynamics. As explained previously, this corresponds to neglecting the linear loss of $\hat{\Psi}_{b_\pm}(z)$, which is valid as long as the linear loss is small during the operation time. In this case, the equations of motion for the field operators are given by
\begin{align}
\partial_t\hat{\mathcal{E}}_{a_+}(z)&=-c\partial_z\hat{\mathcal{E}}_{a_+}(z)+ig_p\hat{\mathcal{P}}_{a_+}(z),\label{eq:eqB29}\\
\partial_t\hat{\mathcal{P}}_{a_+}(z)&=-\gamma\hat{\mathcal{P}}_{a_+}(z)+ig\hat{\mathcal{E}}_{a_+}(z)+i\Omega\hat{\mathcal{S}}_{a_+}(z),\label{eq:eqB30}\\
\partial_t\hat{\mathcal{S}}_{a_+}(z)&=i\Omega\hat{\mathcal{P}}_{a_+}(z)-ig_+^\prime\hat{\Psi}_{b_+}(z)-i\delta\hat{\mathcal{S}}_{a_+}(z)\nonumber\\
&-i\int dz^\prime V(r)\hat{\mathcal{S}}_{a_-}^\dagger(z^\prime)\hat{\mathcal{S}}_{a_-}(z^\prime)\hat{\mathcal{S}}_{a_+}(z),\label{eq:eqB31}\\
\partial_t\hat{\Psi}_{b_+}(z)&=-v\partial_z\hat{\Psi}_{b_+}(z)-ig_+^\prime\hat{\mathcal{S}}_{a_+}(z),\label{eq:eqB32}\\
\partial_t\hat{\mathcal{E}}_{a_-}(z)&=c\partial_z\hat{\mathcal{E}}_{a_-}(z)+ig_p\hat{\mathcal{P}}_{a_-}(z),\label{eq:eqB33}\\
\partial_t\hat{\mathcal{P}}_{a_-}(z)&=-\gamma\hat{\mathcal{P}}_{a_-}(z)+ig\hat{\mathcal{E}}_{a_-}(z)+i\Omega\hat{\mathcal{S}}_{a_-}(z),\label{eq:eqB34}\\
\partial_t\hat{\mathcal{S}}_{a_-}(z)&=i\Omega\hat{\mathcal{P}}_{a_-}(z)-ig_-^\prime\hat{\Psi}_{b_-}(z)-i\delta\hat{\mathcal{S}}_{a_-}(z)\nonumber\\
&-i\int dz^\prime V(r)\hat{\mathcal{S}}_{a_+}^\dagger(z^\prime)\hat{\mathcal{S}}_{a_+}(z^\prime)\hat{\mathcal{S}}_{a_-}(z),\label{eq:eqB35}\\
\partial_t\hat{\Psi}_{b_-}(z)&=v\partial_z\hat{\Psi}_{b_-}(z)-ig_-^\prime\hat{\mathcal{S}}_{a_-}(z),\label{eq:eqB36}
\end{align}
where $g_\pm^\prime=g_\pm\csc\theta_1$, $\delta$ is the two-photon detuning ($\Delta k_\pm=\delta/v$), and $v=c\cos^2\theta$ represents the group velocity of the linear DSP field. Introducing the two-particle wavefunction, the state $|\psi(t)\rangle$ of the system can be described by
\begin{widetext}
\begin{equation}
\psi(z_1,z_2,t)=\left(EE,EP,ES,EB,PE,PP,PS,PB,SA,SP,SS,SB,BE,BP,BS,BB\right)^{T},\label{eq:eqB37}
\end{equation}
with each element being a two-particle wavefunction component (e.g., $BS=\langle0|\hat{\Psi}_{b_+}(z_1)\hat{\mathcal{S}}_{-}(z_2)|\psi(t)\rangle$).
The equation of motion is determined by $i\partial_t\psi(z_1,z_2,t)=\mathcal{H}(z_1,z_2,t)\psi(z_1,z_2,t)$, where the effective Hamiltonian $\mathcal{H}$ is given by
\begin{equation}
\mathcal{H}=
\begin{pmatrix}
-ic\partial_{z_1}&-g_p&0&0\\
-g_p&-i\gamma&-\Omega&0\\
0&-\Omega&\delta&g^\prime_+\\
0&0&g^\prime_+&-iv\partial_{z_1}
\end{pmatrix}\otimes
\begin{pmatrix}
1&0&0&0\\
0&1&0&0\\
0&0&1&0\\
0&0&0&1
\end{pmatrix}+
\begin{pmatrix}
1&0&0&0\\
0&1&0&0\\
0&0&1&0\\
0&0&0&1
\end{pmatrix}\otimes
\begin{pmatrix}
ic\partial_{z_2}&-g_p&0&0\\
-g_p&-i\gamma&-\Omega&0\\
0&-\Omega&\delta&g^\prime_-\\
0&0&g^\prime_-&iv\partial_{z_2}
\end{pmatrix}+
V(z_1-z_2)\Sigma^\dagger\Sigma,\label{eq:eqB38}
\end{equation}
with $\Sigma=(0,0,0,0,0,0,0,0,0,0,1,0,0,0,0,0)$. All numerical results presented in the main text are based on solving this $16\times16$ partial differential equation set with a Gaussian initial pulse.
\end{widetext}

\begin{figure*}
\centering
\includegraphics[width=0.78\linewidth]{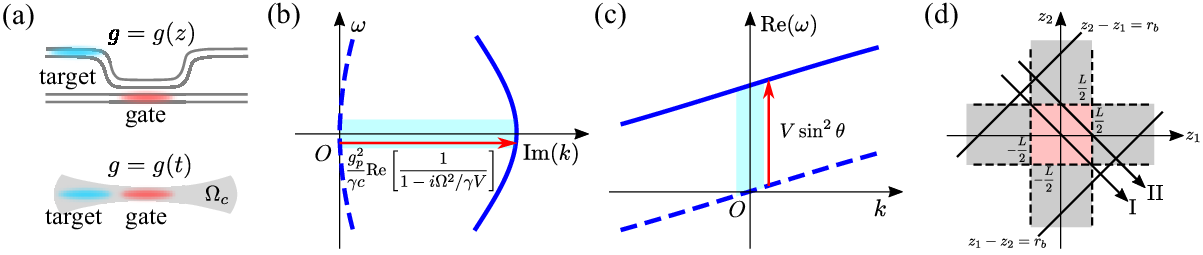}
\caption{(a) Illustration of spatial and temporal modulations of the coupling. (b) Illustration of the interaction induced momentum-mismatch coupling. (c) Illustration of the interaction induced energy-nonresonance coupling. (d) Illustration of the coupling region in the spatial modulation case.}
\label{fig:fig6}
\end{figure*}
\section{Comparison between spatial and temporal modulations of the coupling constant}\label{app:appC}
In this section, we clarify several differences between spatial [$g(z)=g$ for $-L/2<z<L/2$] and temporal [$g(t)=g$ for $-T/2<t<T/2$] modulations of the coupling. Spatial modulation is straightforwardly implemented in the photonic coupling scheme [see the upper panel of Fig.~\ref{fig:fig6}(a)], while temporal modulation can be achieved by varying the intensity of the control field in the atomic coupling scheme [see the lower panel of Fig.~\ref{fig:fig6}(a)]. The temporal modulation of $g$ can also be realized in the photonic coupling scheme by adopting photonic transition approach \cite{yu2009complete} or photonic Raman process \cite{mrejen2015adiabatic}, with the coupling controlled by refraction index modulation.

In fact, for spatial modulation of the coupling constant, the momentum (wave-vector) of the photon is not conserved, while the energy (frequency) of the photon remains conserved. The situation is reversed for temporal modulation. Thus, for spatial modulation, the coupling blockade originates from interaction induced large momentum mismatch [see Fig.~\ref{fig:fig6}(b)], i.e., $|g_p^2/\gamma c(1-i\Omega^2/\gamma V)|\gg g/v$. In this case, perfect blockade requires the compressed length of the stored gate excitation to be smaller than the EIT blockade radius $z_b$ and the coupling region $L<2z_b$. For temporal modulation, coupling blockade occurs if the interaction induced energy shift $V\sin^2\theta$ (in the frequency dimension) is much larger than the coupling $g$ [see Fig.~\ref{fig:fig6}(c)], and perfect blockade requires $\sigma+\frac{1}{2}vT<r_b$ with $\sigma$ the same compressed length of the target and gate photon.

Spatial modulation assisted coupling blockade can also be used for single-photon switching, but is not suitable for applications including entanglement generation or quantum beam splitting unless $\sigma\ll L$. This is due to the nonlocal property of the photon. Taking two counter propagating photons as an example, for the coupling region of length $L$ [as shown in Fig.~\ref{fig:fig6}(d)], the separable two-photon wavefunction is coupled with entangled wavefunction only when $\{(-L/2<z_1<L/2)||(-L/2<z_2<L/2)\}$ (the pink region). For the wavefunction along path I, the initial separable two-photon wavefunction can be efficiently transformed to entangled state, while the wavefunction along path II propagates in regions where only one photon feels the nonzero couplings (gray region) for most of the time and thus entanglement cannot be efficiently generated.

\section{Experimental considerations}\label{app:appD}
In this section, we discuss necessary conditions for the experimental observation of our proposed scheme, mainly concerning by the trade-off between ideal blockade and linear EIT loss.

For DSPs with compressed pulse length $\sigma$ and group velocity $v_g$, the bandwidth of the signal is $\Delta\omega\approx v_g/\sigma$. The loss caused by finite EIT transparency window contains two parts: the loss during the conversion between free space photon and DSPs over the length scale $\sigma$; and the loss during the operation time $g^{-1}$ over the length scale ${v_g/g}$, which are respectively given by
\begin{equation}
\label{eq:eqD1}
\xi = (\Delta\omega)^2 \frac{\gamma g_p^2}{c\Omega^4}\times\sigma,\qquad
\eta = (\Delta\omega)^2 \frac{\gamma g_p^2}{c\Omega^4}\times\frac{v_g}{g}.
\end{equation}
The above equation together with the introduction of coupling blockade radius $r_b$ defined by $C_6/r_b^2=2g$ and EIT blockade radius $z_b$ defined by $C_6/z_b^2=\Omega^2/\gamma$ yield $\sigma = \gamma c/(g_p^2\xi)$ and $r_b = (\eta/2\xi^2)^{1/6} z_b$, which gives
\begin{equation}
\label{eq:eqD2}
r_b/\sigma = (\xi)^{2/3}(\eta/2)^{1/6}\times d_b,
\end{equation}
where $d_b=g_p^2/(\gamma c)z_b$ characterizes the usual blockade optical depth \cite{gorshkov2011photon}.

As mentioned in the main text, ideal coupling blockade requires $r_b>\sigma$. Thus, for a fixed blockade optical depth $d_b$, Eq.~(\ref{eq:eqD2}) indicates that there is a trade-off between the ratio $r_b/\sigma$ and the linear EIT loss $\xi+\eta$, i.e., less loss would lead to worse blockade. In fact, a large optical depth $d_b$ can reduce the loss to acceptable values while keeping the coupling blockade nearly perfect, e.g., for $\xi=5\%$ and $\eta=3\%$, $d_b=15$ can ensure $r_b>\sigma$.

\begin{figure}
\includegraphics[width=\linewidth]{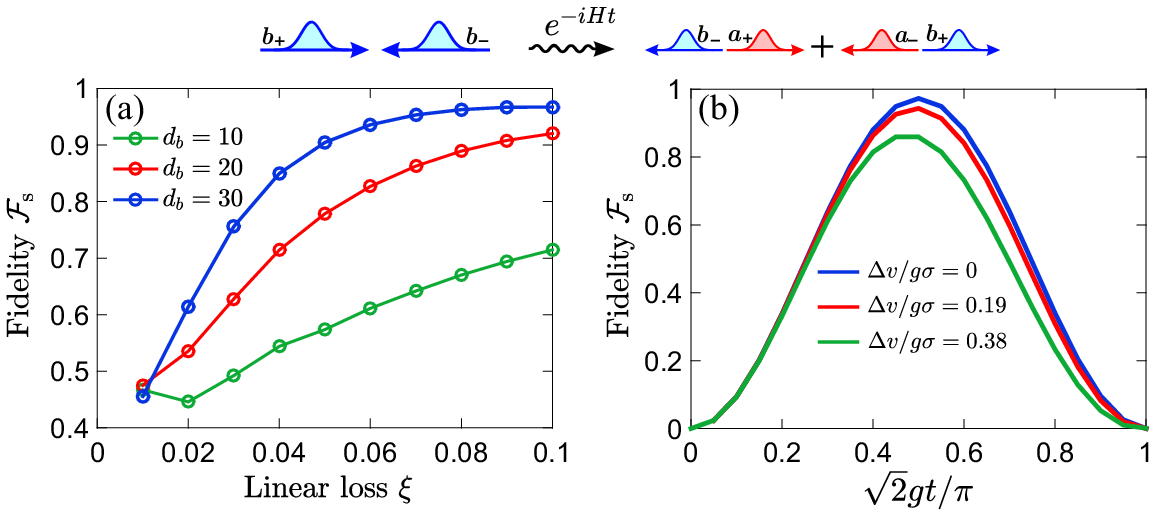}
\caption{(a) Fidelity of the generated symmetric entangled state versus linear EIT loss for the indicated value of blockade optical depth. (b) Oscillation of the fidelity $\mathcal{F}_\mathrm{s}$ for different group velocity mismatches.}
\label{fig:fig7}
\end{figure}

Restricting $\eta$ to be $2.5\%$, we plot the operation fidelity $\mathcal{F}_\mathrm{s}$ of the produced entangled state (see Sec.~\ref{sec:sec4B}) versus the linear EIT loss $\xi$ for different blockade optical depth. As shown in Fig.~\ref{fig:fig7}(a), the increase of the linear EIT loss results in a better operation fidelity (the fidelity during the operation process, not the total fidelity), and a higher $d_b$ yields a higher $\mathcal{F}_\mathrm{s}$ for a fixed linear EIT loss.  Thus, to achieve a high total fidelity, a higher blockade optical depth is preferred. The increase of blockade optical depth $d_b$ can be achieved by using higher-lying Rydberg states, increasing the atomic density, and reducing the transverse area of the photonic mode. In the main text, we use a large coupling constant $g_p=20~\mathrm{GHz}$, corresponding to $d_b=38.46$. Such a strong atom-photon coupling can be achieved by using a Bose-Einstein condenstate with an atomic density $\mathcal{N}\sim10^{14}\mathrm{cm}^{-3}$.

Another requirement arises from the finite life time $\tau_r$ of the Rydberg state. To ensure nearly lossless Rydberg excitation, the time duration for propagation inside the atomic ensemble should be much smaller than $\tau_r$, i.e., $t\sim\sigma/v_g+1/g=(1/\xi+\eta/\xi^2)\gamma/\Omega^2\ll\tau_r$. For $\gamma/2\pi=3$~MHz and $\Omega/2\pi=5$~MHz used in the main text, $\xi=5\%$ and $\eta=3\%$ gives $t\approx0.6~\mu\mathrm{s}$, which meets the requirement as it is shorter than the typical life time $\tau_r$ of the Rydberg polariton (usually up to several microseconds \cite{murray2018photon}).

We also consider the influence of the group velocity mismatch on the blockade effect. The group velocity mismatch between the dark state polariton $\hat{\Psi}_{a_\pm}(z)$ [group velocity being $v_g=c\Omega^2/(g_p^2+\Omega^2)$] and linear light polariton $\hat{\Psi}_{b_\pm}(z)$ (group velocity being $v$) is defined as $\Delta v = v - v_g$. Intuitively, the influence of the group velocity mismatch is determined by the quantity $\Delta vt/\sigma$, i.e., a larger $\Delta vt/\sigma$ suggests a larger separation of the photonic wavefunction in different modes. We plot $\mathcal{F}_\mathrm{s}$ for different values of $v$ [other parameters are the same as in Fig.~\ref{fig:fig2}(c) in the main text] in Fig.~\ref{fig:fig7}(b). If $\Delta v/g\sigma\ll1$ is satisfied, the oscillation visibility remains high during the operation time $t\sim1/g$.

For the single-photon switching in the photonic coupling scheme (spatial modulation case), we only need the stored gate photon to be compressed ($\sigma<z_b$), and the bandwidth of the target photon can be chosen well within the EIT transparency window. In this case, choosing $L=2z_b$ and $gL/v=\pi/2$, the switching fidelity is approximately given by $\mathcal{F}_\mathrm{switch}\approx\exp(-2g^2\gamma c/g_p^2v^2L)=\exp(-\pi^2/4d_b)\approx 1-\pi^2/4d_b$. Such a quantum switching fidelity has the same scaling on $d_b$ as in the novel polariton switching scheme recently proposed \cite{murray2017coherent}. Although the fidelity of our scheme is slightly smaller than \cite{murray2017coherent} in the small $d_b$ region, our scheme possesses a larger bandwidth in the large $d_b$ region, as the bandwidth of the target photon is not reduced by the strong interaction in the coupling blockade scheme.

For the atomic coupling scheme, we provide here a possible choice of level structures for $^{87}\mathrm{Rb}$ atoms. In the scheme, one can choose $|e_+\rangle = |5P_{3/2}, F =2, m_F = 1\rangle$, $|e_-\rangle = |5P_{3/2}, F = 2, m_F = -1\rangle$, $|g\rangle = |5S_{1/2}, F =1, m_F = 0\rangle$, $|d\rangle = |5S_{1/2}, F =2, m_F = 0\rangle$, and $|r\rangle = |nS_{1/2}, J =1/2, m_J = -1/2\rangle$. The coupling $\Omega_c$ between $|d\rangle$ and $|r\rangle$ can be constructed using a two-photon process with an intermediate state $|5P_{1/2},F =1,m_F =-1\rangle$. The $z$-direction wave-vectors $k_1$, $k_2$, and $k_c$ of the control field $\Omega_1$, $\Omega_2$, and $\Omega_c$ should be tuned to satisfy the phase-matching condition $k_1+k_2-k_c=k_{\circlearrowright}-k_{\circlearrowleft}$.

\end{appendix}
\bibliography{manuscript}

\begin{thebibliography}{43}%
\makeatletter
\providecommand \@ifxundefined [1]{%
 \@ifx{#1\undefined}
}%
\providecommand \@ifnum [1]{%
 \ifnum #1\expandafter \@firstoftwo
 \else \expandafter \@secondoftwo
 \fi
}%
\providecommand \@ifx [1]{%
 \ifx #1\expandafter \@firstoftwo
 \else \expandafter \@secondoftwo
 \fi
}%
\providecommand \natexlab [1]{#1}%
\providecommand \enquote  [1]{``#1''}%
\providecommand \bibnamefont  [1]{#1}%
\providecommand \bibfnamefont [1]{#1}%
\providecommand \citenamefont [1]{#1}%
\providecommand \href@noop [0]{\@secondoftwo}%
\providecommand \href [0]{\begingroup \@sanitize@url \@href}%
\providecommand \@href[1]{\@@startlink{#1}\@@href}%
\providecommand \@@href[1]{\endgroup#1\@@endlink}%
\providecommand \@sanitize@url [0]{\catcode `\\12\catcode `\$12\catcode
  `\&12\catcode `\#12\catcode `\^12\catcode `\_12\catcode `\%12\relax}%
\providecommand \@@startlink[1]{}%
\providecommand \@@endlink[0]{}%
\providecommand \url  [0]{\begingroup\@sanitize@url \@url }%
\providecommand \@url [1]{\endgroup\@href {#1}{\urlprefix }}%
\providecommand \urlprefix  [0]{URL }%
\providecommand \Eprint [0]{\href }%
\providecommand \doibase [0]{http://dx.doi.org/}%
\providecommand \selectlanguage [0]{\@gobble}%
\providecommand \bibinfo  [0]{\@secondoftwo}%
\providecommand \bibfield  [0]{\@secondoftwo}%
\providecommand \translation [1]{[#1]}%
\providecommand \BibitemOpen [0]{}%
\providecommand \bibitemStop [0]{}%
\providecommand \bibitemNoStop [0]{.\EOS\space}%
\providecommand \EOS [0]{\spacefactor3000\relax}%
\providecommand \BibitemShut  [1]{\csname bibitem#1\endcsname}%
\let\auto@bib@innerbib\@empty
\bibitem [{\citenamefont {Chang}\ \emph {et~al.}(2014)\citenamefont {Chang},
  \citenamefont {Vuleti{\'c}},\ and\ \citenamefont {Lukin}}]{chang2014quantum}%
  \BibitemOpen
  \bibfield  {author} {\bibinfo {author} {\bibfnamefont {D.~E.}\ \bibnamefont
  {Chang}}, \bibinfo {author} {\bibfnamefont {V.}~\bibnamefont {Vuleti{\'c}}},
  \ and\ \bibinfo {author} {\bibfnamefont {M.~D.}\ \bibnamefont {Lukin}},\
  }\href@noop {} {\bibfield  {journal} {\bibinfo  {journal} {Nat. Photon.}\
  }\textbf {\bibinfo {volume} {8}},\ \bibinfo {pages} {685} (\bibinfo {year}
  {2014})}\BibitemShut {NoStop}%
\bibitem [{\citenamefont {Murray}\ and\ \citenamefont
  {Pohl}(2016)}]{murray2016chapter}%
  \BibitemOpen
  \bibfield  {author} {\bibinfo {author} {\bibfnamefont {C.}~\bibnamefont
  {Murray}}\ and\ \bibinfo {author} {\bibfnamefont {T.}~\bibnamefont {Pohl}},\
  }\href@noop {} {\bibfield  {journal} {\bibinfo  {journal} {Advances In
  Atomic, Molecular, and Optical Physics}\ }\textbf {\bibinfo {volume} {65}},\
  \bibinfo {pages} {321} (\bibinfo {year} {2016})}\BibitemShut {NoStop}%
\bibitem [{\citenamefont {Firstenberg}\ \emph {et~al.}(2016)\citenamefont
  {Firstenberg}, \citenamefont {Adams},\ and\ \citenamefont
  {Hofferberth}}]{firstenberg2016nonlinear}%
  \BibitemOpen
  \bibfield  {author} {\bibinfo {author} {\bibfnamefont {O.}~\bibnamefont
  {Firstenberg}}, \bibinfo {author} {\bibfnamefont {C.~S.}\ \bibnamefont
  {Adams}}, \ and\ \bibinfo {author} {\bibfnamefont {S.}~\bibnamefont
  {Hofferberth}},\ }\href@noop {} {\bibfield  {journal} {\bibinfo  {journal}
  {J. Phys. B}\ }\textbf {\bibinfo {volume} {49}},\ \bibinfo {pages} {152003}
  (\bibinfo {year} {2016})}\BibitemShut {NoStop}%
\bibitem [{\citenamefont {Friedler}\ \emph {et~al.}(2005)\citenamefont
  {Friedler}, \citenamefont {Petrosyan}, \citenamefont {Fleischhauer},\ and\
  \citenamefont {Kurizki}}]{friedler2005long}%
  \BibitemOpen
  \bibfield  {author} {\bibinfo {author} {\bibfnamefont {I.}~\bibnamefont
  {Friedler}}, \bibinfo {author} {\bibfnamefont {D.}~\bibnamefont {Petrosyan}},
  \bibinfo {author} {\bibfnamefont {M.}~\bibnamefont {Fleischhauer}}, \ and\
  \bibinfo {author} {\bibfnamefont {G.}~\bibnamefont {Kurizki}},\ }\href@noop
  {} {\bibfield  {journal} {\bibinfo  {journal} {Phys. Rev. A}\ }\textbf
  {\bibinfo {volume} {72}},\ \bibinfo {pages} {043803} (\bibinfo {year}
  {2005})}\BibitemShut {NoStop}%
\bibitem [{\citenamefont {Kimble}(2008)}]{kimble2008quantum}%
  \BibitemOpen
  \bibfield  {author} {\bibinfo {author} {\bibfnamefont {H.}~\bibnamefont
  {Kimble}},\ }\href@noop {} {\bibfield  {journal} {\bibinfo  {journal}
  {Nature}\ }\textbf {\bibinfo {volume} {453}},\ \bibinfo {pages} {1023}
  (\bibinfo {year} {2008})}\BibitemShut {NoStop}%
\bibitem [{\citenamefont {O'brien}\ \emph {et~al.}(2009)\citenamefont
  {O'brien}, \citenamefont {Furusawa},\ and\ \citenamefont
  {Vu{\v{c}}kovi{\'c}}}]{o2009photonic}%
  \BibitemOpen
  \bibfield  {author} {\bibinfo {author} {\bibfnamefont {J.~L.}\ \bibnamefont
  {O'brien}}, \bibinfo {author} {\bibfnamefont {A.}~\bibnamefont {Furusawa}}, \
  and\ \bibinfo {author} {\bibfnamefont {J.}~\bibnamefont
  {Vu{\v{c}}kovi{\'c}}},\ }\href@noop {} {\bibfield  {journal} {\bibinfo
  {journal} {Nat. Photon.}\ }\textbf {\bibinfo {volume} {3}},\ \bibinfo {pages}
  {687} (\bibinfo {year} {2009})}\BibitemShut {NoStop}%
\bibitem [{\citenamefont {Chang}\ \emph {et~al.}(2008)\citenamefont {Chang},
  \citenamefont {Gritsev}, \citenamefont {Morigi}, \citenamefont {Vuletic},
  \citenamefont {Lukin},\ and\ \citenamefont
  {Demler}}]{chang2008crystallization}%
  \BibitemOpen
  \bibfield  {author} {\bibinfo {author} {\bibfnamefont {D.}~\bibnamefont
  {Chang}}, \bibinfo {author} {\bibfnamefont {V.}~\bibnamefont {Gritsev}},
  \bibinfo {author} {\bibfnamefont {G.}~\bibnamefont {Morigi}}, \bibinfo
  {author} {\bibfnamefont {V.}~\bibnamefont {Vuletic}}, \bibinfo {author}
  {\bibfnamefont {M.}~\bibnamefont {Lukin}}, \ and\ \bibinfo {author}
  {\bibfnamefont {E.}~\bibnamefont {Demler}},\ }\href@noop {} {\bibfield
  {journal} {\bibinfo  {journal} {Nat. Phys.}\ }\textbf {\bibinfo {volume}
  {4}},\ \bibinfo {pages} {884} (\bibinfo {year} {2008})}\BibitemShut {NoStop}%
\bibitem [{\citenamefont {Otterbach}\ \emph {et~al.}(2013)\citenamefont
  {Otterbach}, \citenamefont {Moos}, \citenamefont {Muth},\ and\ \citenamefont
  {Fleischhauer}}]{otterbach2013wigner}%
  \BibitemOpen
  \bibfield  {author} {\bibinfo {author} {\bibfnamefont {J.}~\bibnamefont
  {Otterbach}}, \bibinfo {author} {\bibfnamefont {M.}~\bibnamefont {Moos}},
  \bibinfo {author} {\bibfnamefont {D.}~\bibnamefont {Muth}}, \ and\ \bibinfo
  {author} {\bibfnamefont {M.}~\bibnamefont {Fleischhauer}},\ }\href@noop {}
  {\bibfield  {journal} {\bibinfo  {journal} {Phys. Rev. Lett.}\ }\textbf
  {\bibinfo {volume} {111}},\ \bibinfo {pages} {113001} (\bibinfo {year}
  {2013})}\BibitemShut {NoStop}%
\bibitem [{\citenamefont {Gullans}\ \emph {et~al.}(2016)\citenamefont
  {Gullans}, \citenamefont {Thompson}, \citenamefont {Wang}, \citenamefont
  {Liang}, \citenamefont {Vuleti{\'c}}, \citenamefont {Lukin},\ and\
  \citenamefont {Gorshkov}}]{gullans2016effective}%
  \BibitemOpen
  \bibfield  {author} {\bibinfo {author} {\bibfnamefont {M.}~\bibnamefont
  {Gullans}}, \bibinfo {author} {\bibfnamefont {J.}~\bibnamefont {Thompson}},
  \bibinfo {author} {\bibfnamefont {Y.}~\bibnamefont {Wang}}, \bibinfo {author}
  {\bibfnamefont {Q.-Y.}\ \bibnamefont {Liang}}, \bibinfo {author}
  {\bibfnamefont {V.}~\bibnamefont {Vuleti{\'c}}}, \bibinfo {author}
  {\bibfnamefont {M.~D.}\ \bibnamefont {Lukin}}, \ and\ \bibinfo {author}
  {\bibfnamefont {A.~V.}\ \bibnamefont {Gorshkov}},\ }\href@noop {} {\bibfield
  {journal} {\bibinfo  {journal} {Phys. Rev. Lett.}\ }\textbf {\bibinfo
  {volume} {117}},\ \bibinfo {pages} {113601} (\bibinfo {year}
  {2016})}\BibitemShut {NoStop}%
\bibitem [{\citenamefont {Murray}\ \emph {et~al.}(2018)\citenamefont {Murray},
  \citenamefont {Mirgorodskiy}, \citenamefont {Tresp}, \citenamefont {Braun},
  \citenamefont {Paris-Mandoki}, \citenamefont {Gorshkov}, \citenamefont
  {Hofferberth},\ and\ \citenamefont {Pohl}}]{murray2018photon}%
  \BibitemOpen
  \bibfield  {author} {\bibinfo {author} {\bibfnamefont {C.~R.}\ \bibnamefont
  {Murray}}, \bibinfo {author} {\bibfnamefont {I.}~\bibnamefont
  {Mirgorodskiy}}, \bibinfo {author} {\bibfnamefont {C.}~\bibnamefont {Tresp}},
  \bibinfo {author} {\bibfnamefont {C.}~\bibnamefont {Braun}}, \bibinfo
  {author} {\bibfnamefont {A.}~\bibnamefont {Paris-Mandoki}}, \bibinfo {author}
  {\bibfnamefont {A.~V.}\ \bibnamefont {Gorshkov}}, \bibinfo {author}
  {\bibfnamefont {S.}~\bibnamefont {Hofferberth}}, \ and\ \bibinfo {author}
  {\bibfnamefont {T.}~\bibnamefont {Pohl}},\ }\href@noop {} {\bibfield
  {journal} {\bibinfo  {journal} {Phys. Rev. Lett.}\ }\textbf {\bibinfo
  {volume} {120}},\ \bibinfo {pages} {113601} (\bibinfo {year}
  {2018})}\BibitemShut {NoStop}%
\bibitem [{\citenamefont {Birnbaum}\ \emph {et~al.}(2005)\citenamefont
  {Birnbaum}, \citenamefont {Boca}, \citenamefont {Miller}, \citenamefont
  {Boozer}, \citenamefont {Northup},\ and\ \citenamefont
  {Kimble}}]{birnbaum2005photon}%
  \BibitemOpen
  \bibfield  {author} {\bibinfo {author} {\bibfnamefont {K.~M.}\ \bibnamefont
  {Birnbaum}}, \bibinfo {author} {\bibfnamefont {A.}~\bibnamefont {Boca}},
  \bibinfo {author} {\bibfnamefont {R.}~\bibnamefont {Miller}}, \bibinfo
  {author} {\bibfnamefont {A.~D.}\ \bibnamefont {Boozer}}, \bibinfo {author}
  {\bibfnamefont {T.~E.}\ \bibnamefont {Northup}}, \ and\ \bibinfo {author}
  {\bibfnamefont {H.~J.}\ \bibnamefont {Kimble}},\ }\href@noop {} {\bibfield
  {journal} {\bibinfo  {journal} {Nature}\ }\textbf {\bibinfo {volume} {436}},\
  \bibinfo {pages} {87} (\bibinfo {year} {2005})}\BibitemShut {NoStop}%
\bibitem [{\citenamefont {Thompson}\ \emph {et~al.}(2013)\citenamefont
  {Thompson}, \citenamefont {Tiecke}, \citenamefont {de~Leon}, \citenamefont
  {Feist}, \citenamefont {Akimov}, \citenamefont {Gullans}, \citenamefont
  {Zibrov}, \citenamefont {Vuleti{\'c}},\ and\ \citenamefont
  {Lukin}}]{thompson2013coupling}%
  \BibitemOpen
  \bibfield  {author} {\bibinfo {author} {\bibfnamefont {J.~D.}\ \bibnamefont
  {Thompson}}, \bibinfo {author} {\bibfnamefont {T.}~\bibnamefont {Tiecke}},
  \bibinfo {author} {\bibfnamefont {N.~P.}\ \bibnamefont {de~Leon}}, \bibinfo
  {author} {\bibfnamefont {J.}~\bibnamefont {Feist}}, \bibinfo {author}
  {\bibfnamefont {A.}~\bibnamefont {Akimov}}, \bibinfo {author} {\bibfnamefont
  {M.}~\bibnamefont {Gullans}}, \bibinfo {author} {\bibfnamefont {A.~S.}\
  \bibnamefont {Zibrov}}, \bibinfo {author} {\bibfnamefont {V.}~\bibnamefont
  {Vuleti{\'c}}}, \ and\ \bibinfo {author} {\bibfnamefont {M.~D.}\ \bibnamefont
  {Lukin}},\ }\href@noop {} {\bibfield  {journal} {\bibinfo  {journal}
  {Science}\ }\textbf {\bibinfo {volume} {340}},\ \bibinfo {pages} {1202}
  (\bibinfo {year} {2013})}\BibitemShut {NoStop}%
\bibitem [{\citenamefont {Goban}\ \emph {et~al.}(2014)\citenamefont {Goban},
  \citenamefont {Hung}, \citenamefont {Yu}, \citenamefont {Hood}, \citenamefont
  {Muniz}, \citenamefont {Lee}, \citenamefont {Martin}, \citenamefont
  {McClung}, \citenamefont {Choi}, \citenamefont {Chang} \emph
  {et~al.}}]{goban2014atom}%
  \BibitemOpen
  \bibfield  {author} {\bibinfo {author} {\bibfnamefont {A.}~\bibnamefont
  {Goban}}, \bibinfo {author} {\bibfnamefont {C.-L.}\ \bibnamefont {Hung}},
  \bibinfo {author} {\bibfnamefont {S.-P.}\ \bibnamefont {Yu}}, \bibinfo
  {author} {\bibfnamefont {J.}~\bibnamefont {Hood}}, \bibinfo {author}
  {\bibfnamefont {J.}~\bibnamefont {Muniz}}, \bibinfo {author} {\bibfnamefont
  {J.}~\bibnamefont {Lee}}, \bibinfo {author} {\bibfnamefont {M.}~\bibnamefont
  {Martin}}, \bibinfo {author} {\bibfnamefont {A.}~\bibnamefont {McClung}},
  \bibinfo {author} {\bibfnamefont {K.}~\bibnamefont {Choi}}, \bibinfo {author}
  {\bibfnamefont {D.~E.}\ \bibnamefont {Chang}},  \emph {et~al.},\ }\href@noop
  {} {\bibfield  {journal} {\bibinfo  {journal} {Nat. Commun.}\ }\textbf
  {\bibinfo {volume} {5}} (\bibinfo {year} {2014})}\BibitemShut {NoStop}%
\bibitem [{\citenamefont {Reiserer}\ and\ \citenamefont
  {Rempe}(2015)}]{reiserer2015cavity}%
  \BibitemOpen
  \bibfield  {author} {\bibinfo {author} {\bibfnamefont {A.}~\bibnamefont
  {Reiserer}}\ and\ \bibinfo {author} {\bibfnamefont {G.}~\bibnamefont
  {Rempe}},\ }\href@noop {} {\bibfield  {journal} {\bibinfo  {journal} {Rev.
  Mod. Phys.}\ }\textbf {\bibinfo {volume} {87}},\ \bibinfo {pages} {1379}
  (\bibinfo {year} {2015})}\BibitemShut {NoStop}%
\bibitem [{\citenamefont {Douglas}\ \emph {et~al.}(2016)\citenamefont
  {Douglas}, \citenamefont {Caneva},\ and\ \citenamefont
  {Chang}}]{douglas2016photon}%
  \BibitemOpen
  \bibfield  {author} {\bibinfo {author} {\bibfnamefont {J.~S.}\ \bibnamefont
  {Douglas}}, \bibinfo {author} {\bibfnamefont {T.}~\bibnamefont {Caneva}}, \
  and\ \bibinfo {author} {\bibfnamefont {D.~E.}\ \bibnamefont {Chang}},\
  }\href@noop {} {\bibfield  {journal} {\bibinfo  {journal} {Phys. Rev. X}\
  }\textbf {\bibinfo {volume} {6}},\ \bibinfo {pages} {031017} (\bibinfo {year}
  {2016})}\BibitemShut {NoStop}%
\bibitem [{\citenamefont {Shahmoon}\ \emph {et~al.}(2016)\citenamefont
  {Shahmoon}, \citenamefont {Gri{\v{s}}ins}, \citenamefont {Stimming},
  \citenamefont {Mazets},\ and\ \citenamefont {Kurizki}}]{shahmoon2016highly}%
  \BibitemOpen
  \bibfield  {author} {\bibinfo {author} {\bibfnamefont {E.}~\bibnamefont
  {Shahmoon}}, \bibinfo {author} {\bibfnamefont {P.}~\bibnamefont
  {Gri{\v{s}}ins}}, \bibinfo {author} {\bibfnamefont {H.~P.}\ \bibnamefont
  {Stimming}}, \bibinfo {author} {\bibfnamefont {I.}~\bibnamefont {Mazets}}, \
  and\ \bibinfo {author} {\bibfnamefont {G.}~\bibnamefont {Kurizki}},\
  }\href@noop {} {\bibfield  {journal} {\bibinfo  {journal} {Optica}\ }\textbf
  {\bibinfo {volume} {3}},\ \bibinfo {pages} {725} (\bibinfo {year}
  {2016})}\BibitemShut {NoStop}%
\bibitem [{\citenamefont {Gorshkov}\ \emph {et~al.}(2011)\citenamefont
  {Gorshkov}, \citenamefont {Otterbach}, \citenamefont {Fleischhauer},
  \citenamefont {Pohl},\ and\ \citenamefont {Lukin}}]{gorshkov2011photon}%
  \BibitemOpen
  \bibfield  {author} {\bibinfo {author} {\bibfnamefont {A.~V.}\ \bibnamefont
  {Gorshkov}}, \bibinfo {author} {\bibfnamefont {J.}~\bibnamefont {Otterbach}},
  \bibinfo {author} {\bibfnamefont {M.}~\bibnamefont {Fleischhauer}}, \bibinfo
  {author} {\bibfnamefont {T.}~\bibnamefont {Pohl}}, \ and\ \bibinfo {author}
  {\bibfnamefont {M.~D.}\ \bibnamefont {Lukin}},\ }\href@noop {} {\bibfield
  {journal} {\bibinfo  {journal} {Phys. Rev. Lett.}\ }\textbf {\bibinfo
  {volume} {107}},\ \bibinfo {pages} {133602} (\bibinfo {year}
  {2011})}\BibitemShut {NoStop}%
\bibitem [{\citenamefont {Maghrebi}\ \emph {et~al.}(2015)\citenamefont
  {Maghrebi}, \citenamefont {Gullans}, \citenamefont {Bienias}, \citenamefont
  {Choi}, \citenamefont {Martin}, \citenamefont {Firstenberg}, \citenamefont
  {Lukin}, \citenamefont {B{\"u}chler},\ and\ \citenamefont
  {Gorshkov}}]{maghrebi2015coulomb}%
  \BibitemOpen
  \bibfield  {author} {\bibinfo {author} {\bibfnamefont {M.~F.}\ \bibnamefont
  {Maghrebi}}, \bibinfo {author} {\bibfnamefont {M.~J.}\ \bibnamefont
  {Gullans}}, \bibinfo {author} {\bibfnamefont {P.}~\bibnamefont {Bienias}},
  \bibinfo {author} {\bibfnamefont {S.}~\bibnamefont {Choi}}, \bibinfo {author}
  {\bibfnamefont {I.}~\bibnamefont {Martin}}, \bibinfo {author} {\bibfnamefont
  {O.}~\bibnamefont {Firstenberg}}, \bibinfo {author} {\bibfnamefont {M.~D.}\
  \bibnamefont {Lukin}}, \bibinfo {author} {\bibfnamefont {H.}~\bibnamefont
  {B{\"u}chler}}, \ and\ \bibinfo {author} {\bibfnamefont {A.~V.}\ \bibnamefont
  {Gorshkov}},\ }\href@noop {} {\bibfield  {journal} {\bibinfo  {journal}
  {Phys. Rev. Lett.}\ }\textbf {\bibinfo {volume} {115}},\ \bibinfo {pages}
  {123601} (\bibinfo {year} {2015})}\BibitemShut {NoStop}%
\bibitem [{\citenamefont {Peyronel}\ \emph {et~al.}(2012)\citenamefont
  {Peyronel}, \citenamefont {Firstenberg}, \citenamefont {Liang}, \citenamefont
  {Hofferberth}, \citenamefont {Gorshkov}, \citenamefont {Pohl}, \citenamefont
  {Lukin},\ and\ \citenamefont {Vuletic}}]{peyronel2012quantum}%
  \BibitemOpen
  \bibfield  {author} {\bibinfo {author} {\bibfnamefont {T.}~\bibnamefont
  {Peyronel}}, \bibinfo {author} {\bibfnamefont {O.}~\bibnamefont
  {Firstenberg}}, \bibinfo {author} {\bibfnamefont {Q.-Y.}\ \bibnamefont
  {Liang}}, \bibinfo {author} {\bibfnamefont {S.}~\bibnamefont {Hofferberth}},
  \bibinfo {author} {\bibfnamefont {A.~V.}\ \bibnamefont {Gorshkov}}, \bibinfo
  {author} {\bibfnamefont {T.}~\bibnamefont {Pohl}}, \bibinfo {author}
  {\bibfnamefont {M.~D.}\ \bibnamefont {Lukin}}, \ and\ \bibinfo {author}
  {\bibfnamefont {V.}~\bibnamefont {Vuletic}},\ }\href@noop {} {\bibfield
  {journal} {\bibinfo  {journal} {Nature}\ }\textbf {\bibinfo {volume} {487}},\
  \bibinfo {pages} {57} (\bibinfo {year} {2012})}\BibitemShut {NoStop}%
\bibitem [{\citenamefont {Firstenberg}\ \emph {et~al.}(2013)\citenamefont
  {Firstenberg}, \citenamefont {Peyronel}, \citenamefont {Liang}, \citenamefont
  {Gorshkov}, \citenamefont {Lukin},\ and\ \citenamefont
  {Vuleti{\'c}}}]{firstenberg2013attractive}%
  \BibitemOpen
  \bibfield  {author} {\bibinfo {author} {\bibfnamefont {O.}~\bibnamefont
  {Firstenberg}}, \bibinfo {author} {\bibfnamefont {T.}~\bibnamefont
  {Peyronel}}, \bibinfo {author} {\bibfnamefont {Q.-Y.}\ \bibnamefont {Liang}},
  \bibinfo {author} {\bibfnamefont {A.~V.}\ \bibnamefont {Gorshkov}}, \bibinfo
  {author} {\bibfnamefont {M.~D.}\ \bibnamefont {Lukin}}, \ and\ \bibinfo
  {author} {\bibfnamefont {V.}~\bibnamefont {Vuleti{\'c}}},\ }\href@noop {}
  {\bibfield  {journal} {\bibinfo  {journal} {Nature}\ }\textbf {\bibinfo
  {volume} {502}},\ \bibinfo {pages} {71} (\bibinfo {year} {2013})}\BibitemShut
  {NoStop}%
\bibitem [{\citenamefont {Baur}\ \emph {et~al.}(2014)\citenamefont {Baur},
  \citenamefont {Tiarks}, \citenamefont {Rempe},\ and\ \citenamefont
  {D{\"u}rr}}]{baur2014single}%
  \BibitemOpen
  \bibfield  {author} {\bibinfo {author} {\bibfnamefont {S.}~\bibnamefont
  {Baur}}, \bibinfo {author} {\bibfnamefont {D.}~\bibnamefont {Tiarks}},
  \bibinfo {author} {\bibfnamefont {G.}~\bibnamefont {Rempe}}, \ and\ \bibinfo
  {author} {\bibfnamefont {S.}~\bibnamefont {D{\"u}rr}},\ }\href@noop {}
  {\bibfield  {journal} {\bibinfo  {journal} {Phys. Rev. Lett.}\ }\textbf
  {\bibinfo {volume} {112}},\ \bibinfo {pages} {073901} (\bibinfo {year}
  {2014})}\BibitemShut {NoStop}%
\bibitem [{\citenamefont {Gorniaczyk}\ \emph {et~al.}(2014)\citenamefont
  {Gorniaczyk}, \citenamefont {Tresp}, \citenamefont {Schmidt}, \citenamefont
  {Fedder},\ and\ \citenamefont {Hofferberth}}]{gorniaczyk2014single}%
  \BibitemOpen
  \bibfield  {author} {\bibinfo {author} {\bibfnamefont {H.}~\bibnamefont
  {Gorniaczyk}}, \bibinfo {author} {\bibfnamefont {C.}~\bibnamefont {Tresp}},
  \bibinfo {author} {\bibfnamefont {J.}~\bibnamefont {Schmidt}}, \bibinfo
  {author} {\bibfnamefont {H.}~\bibnamefont {Fedder}}, \ and\ \bibinfo {author}
  {\bibfnamefont {S.}~\bibnamefont {Hofferberth}},\ }\href@noop {} {\bibfield
  {journal} {\bibinfo  {journal} {Phys. Rev. Lett.}\ }\textbf {\bibinfo
  {volume} {113}},\ \bibinfo {pages} {053601} (\bibinfo {year}
  {2014})}\BibitemShut {NoStop}%
\bibitem [{\citenamefont {Tresp}\ \emph {et~al.}(2016)\citenamefont {Tresp},
  \citenamefont {Zimmer}, \citenamefont {Mirgorodskiy}, \citenamefont
  {Gorniaczyk}, \citenamefont {Paris-Mandoki},\ and\ \citenamefont
  {Hofferberth}}]{tresp2016single}%
  \BibitemOpen
  \bibfield  {author} {\bibinfo {author} {\bibfnamefont {C.}~\bibnamefont
  {Tresp}}, \bibinfo {author} {\bibfnamefont {C.}~\bibnamefont {Zimmer}},
  \bibinfo {author} {\bibfnamefont {I.}~\bibnamefont {Mirgorodskiy}}, \bibinfo
  {author} {\bibfnamefont {H.}~\bibnamefont {Gorniaczyk}}, \bibinfo {author}
  {\bibfnamefont {A.}~\bibnamefont {Paris-Mandoki}}, \ and\ \bibinfo {author}
  {\bibfnamefont {S.}~\bibnamefont {Hofferberth}},\ }\href@noop {} {\bibfield
  {journal} {\bibinfo  {journal} {Phys. Rev. Lett.}\ }\textbf {\bibinfo
  {volume} {117}},\ \bibinfo {pages} {223001} (\bibinfo {year}
  {2016})}\BibitemShut {NoStop}%
\bibitem [{\citenamefont {Distante}\ \emph {et~al.}(2016)\citenamefont
  {Distante}, \citenamefont {Padr{\'o}n-Brito}, \citenamefont {Cristiani},
  \citenamefont {Paredes-Barato},\ and\ \citenamefont
  {de~Riedmatten}}]{distante2016storage}%
  \BibitemOpen
  \bibfield  {author} {\bibinfo {author} {\bibfnamefont {E.}~\bibnamefont
  {Distante}}, \bibinfo {author} {\bibfnamefont {A.}~\bibnamefont
  {Padr{\'o}n-Brito}}, \bibinfo {author} {\bibfnamefont {M.}~\bibnamefont
  {Cristiani}}, \bibinfo {author} {\bibfnamefont {D.}~\bibnamefont
  {Paredes-Barato}}, \ and\ \bibinfo {author} {\bibfnamefont {H.}~\bibnamefont
  {de~Riedmatten}},\ }\href@noop {} {\bibfield  {journal} {\bibinfo  {journal}
  {Phys. Rev. Lett.}\ }\textbf {\bibinfo {volume} {117}},\ \bibinfo {pages}
  {113001} (\bibinfo {year} {2016})}\BibitemShut {NoStop}%
\bibitem [{\citenamefont {Gorshkov}\ \emph {et~al.}(2013)\citenamefont
  {Gorshkov}, \citenamefont {Nath},\ and\ \citenamefont
  {Pohl}}]{gorshkov2013dissipative}%
  \BibitemOpen
  \bibfield  {author} {\bibinfo {author} {\bibfnamefont {A.~V.}\ \bibnamefont
  {Gorshkov}}, \bibinfo {author} {\bibfnamefont {R.}~\bibnamefont {Nath}}, \
  and\ \bibinfo {author} {\bibfnamefont {T.}~\bibnamefont {Pohl}},\ }\href@noop
  {} {\bibfield  {journal} {\bibinfo  {journal} {Phys. Rev. Lett.}\ }\textbf
  {\bibinfo {volume} {110}},\ \bibinfo {pages} {153601} (\bibinfo {year}
  {2013})}\BibitemShut {NoStop}%
\bibitem [{\citenamefont {Murray}\ \emph {et~al.}(2016)\citenamefont {Murray},
  \citenamefont {Gorshkov},\ and\ \citenamefont {Pohl}}]{murray2016many}%
  \BibitemOpen
  \bibfield  {author} {\bibinfo {author} {\bibfnamefont {C.~R.}\ \bibnamefont
  {Murray}}, \bibinfo {author} {\bibfnamefont {A.~V.}\ \bibnamefont
  {Gorshkov}}, \ and\ \bibinfo {author} {\bibfnamefont {T.}~\bibnamefont
  {Pohl}},\ }\href@noop {} {\bibfield  {journal} {\bibinfo  {journal} {New J.
  Phys.}\ }\textbf {\bibinfo {volume} {18}},\ \bibinfo {pages} {092001}
  (\bibinfo {year} {2016})}\BibitemShut {NoStop}%
\bibitem [{\citenamefont {Murray}\ and\ \citenamefont
  {Pohl}(2017)}]{murray2017coherent}%
  \BibitemOpen
  \bibfield  {author} {\bibinfo {author} {\bibfnamefont {C.~R.}\ \bibnamefont
  {Murray}}\ and\ \bibinfo {author} {\bibfnamefont {T.}~\bibnamefont {Pohl}},\
  }\href@noop {} {\bibfield  {journal} {\bibinfo  {journal} {Phys. Rev. X}\
  }\textbf {\bibinfo {volume} {7}},\ \bibinfo {pages} {031007} (\bibinfo {year}
  {2017})}\BibitemShut {NoStop}%
\bibitem [{\citenamefont {Yariv}\ and\ \citenamefont
  {Yeh}(2007)}]{yariv2007photonics}%
  \BibitemOpen
  \bibfield  {author} {\bibinfo {author} {\bibfnamefont {A.}~\bibnamefont
  {Yariv}}\ and\ \bibinfo {author} {\bibfnamefont {P.}~\bibnamefont {Yeh}},\
  }\href@noop {} {\emph {\bibinfo {title} {Photonics: optical electronics in
  modern communications}}},\ Vol.~\bibinfo {volume} {6}\ (\bibinfo  {publisher}
  {Oxford University Press New York},\ \bibinfo {year} {2007})\BibitemShut
  {NoStop}%
\bibitem [{\citenamefont {Lukin}\ \emph {et~al.}(2001)\citenamefont {Lukin},
  \citenamefont {Fleischhauer}, \citenamefont {Cote}, \citenamefont {Duan},
  \citenamefont {Jaksch}, \citenamefont {Cirac},\ and\ \citenamefont
  {Zoller}}]{lukin2001dipole}%
  \BibitemOpen
  \bibfield  {author} {\bibinfo {author} {\bibfnamefont {M.}~\bibnamefont
  {Lukin}}, \bibinfo {author} {\bibfnamefont {M.}~\bibnamefont {Fleischhauer}},
  \bibinfo {author} {\bibfnamefont {R.}~\bibnamefont {Cote}}, \bibinfo {author}
  {\bibfnamefont {L.}~\bibnamefont {Duan}}, \bibinfo {author} {\bibfnamefont
  {D.}~\bibnamefont {Jaksch}}, \bibinfo {author} {\bibfnamefont
  {J.}~\bibnamefont {Cirac}}, \ and\ \bibinfo {author} {\bibfnamefont
  {P.}~\bibnamefont {Zoller}},\ }\href@noop {} {\bibfield  {journal} {\bibinfo
  {journal} {Phys. Rev. Lett.}\ }\textbf {\bibinfo {volume} {87}},\ \bibinfo
  {pages} {037901} (\bibinfo {year} {2001})}\BibitemShut {NoStop}%
\bibitem [{\citenamefont {Urban}\ \emph {et~al.}(2009)\citenamefont {Urban},
  \citenamefont {Johnson}, \citenamefont {Henage}, \citenamefont {Isenhower},
  \citenamefont {Yavuz}, \citenamefont {Walker},\ and\ \citenamefont
  {Saffman}}]{urban2009observation}%
  \BibitemOpen
  \bibfield  {author} {\bibinfo {author} {\bibfnamefont {E.}~\bibnamefont
  {Urban}}, \bibinfo {author} {\bibfnamefont {T.}~\bibnamefont {Johnson}},
  \bibinfo {author} {\bibfnamefont {T.}~\bibnamefont {Henage}}, \bibinfo
  {author} {\bibfnamefont {L.}~\bibnamefont {Isenhower}}, \bibinfo {author}
  {\bibfnamefont {D.}~\bibnamefont {Yavuz}}, \bibinfo {author} {\bibfnamefont
  {T.}~\bibnamefont {Walker}}, \ and\ \bibinfo {author} {\bibfnamefont
  {M.}~\bibnamefont {Saffman}},\ }\href@noop {} {\bibfield  {journal} {\bibinfo
   {journal} {Nat. Phys.}\ }\textbf {\bibinfo {volume} {5}},\ \bibinfo {pages}
  {110} (\bibinfo {year} {2009})}\BibitemShut {NoStop}%
\bibitem [{\citenamefont {Gaetan}\ \emph {et~al.}(2009)\citenamefont {Gaetan},
  \citenamefont {Miroshnychenko}, \citenamefont {Wilk}, \citenamefont {Chotia},
  \citenamefont {Viteau}, \citenamefont {Comparat}, \citenamefont {Pillet},
  \citenamefont {Browaeys},\ and\ \citenamefont
  {Grangier}}]{gaetan2009observation}%
  \BibitemOpen
  \bibfield  {author} {\bibinfo {author} {\bibfnamefont {A.}~\bibnamefont
  {Gaetan}}, \bibinfo {author} {\bibfnamefont {Y.}~\bibnamefont
  {Miroshnychenko}}, \bibinfo {author} {\bibfnamefont {T.}~\bibnamefont
  {Wilk}}, \bibinfo {author} {\bibfnamefont {A.}~\bibnamefont {Chotia}},
  \bibinfo {author} {\bibfnamefont {M.}~\bibnamefont {Viteau}}, \bibinfo
  {author} {\bibfnamefont {D.}~\bibnamefont {Comparat}}, \bibinfo {author}
  {\bibfnamefont {P.}~\bibnamefont {Pillet}}, \bibinfo {author} {\bibfnamefont
  {A.}~\bibnamefont {Browaeys}}, \ and\ \bibinfo {author} {\bibfnamefont
  {P.}~\bibnamefont {Grangier}},\ }\href@noop {} {\bibfield  {journal}
  {\bibinfo  {journal} {Nat. Phys.}\ }\textbf {\bibinfo {volume} {5}},\
  \bibinfo {pages} {115} (\bibinfo {year} {2009})}\BibitemShut {NoStop}%
\bibitem [{\citenamefont {Fleischhauer}\ and\ \citenamefont
  {Lukin}(2000)}]{fleischhauer2000dark}%
  \BibitemOpen
  \bibfield  {author} {\bibinfo {author} {\bibfnamefont {M.}~\bibnamefont
  {Fleischhauer}}\ and\ \bibinfo {author} {\bibfnamefont {M.~D.}\ \bibnamefont
  {Lukin}},\ }\href@noop {} {\bibfield  {journal} {\bibinfo  {journal} {Phys.
  Rev. Lett.}\ }\textbf {\bibinfo {volume} {84}},\ \bibinfo {pages} {5094}
  (\bibinfo {year} {2000})}\BibitemShut {NoStop}%
\bibitem [{\citenamefont {Christensen}\ \emph {et~al.}(2008)\citenamefont
  {Christensen}, \citenamefont {Will}, \citenamefont {Saba}, \citenamefont
  {Jo}, \citenamefont {Shin}, \citenamefont {Ketterle},\ and\ \citenamefont
  {Pritchard}}]{christensen2008trapping}%
  \BibitemOpen
  \bibfield  {author} {\bibinfo {author} {\bibfnamefont {C.~A.}\ \bibnamefont
  {Christensen}}, \bibinfo {author} {\bibfnamefont {S.}~\bibnamefont {Will}},
  \bibinfo {author} {\bibfnamefont {M.}~\bibnamefont {Saba}}, \bibinfo {author}
  {\bibfnamefont {G.-B.}\ \bibnamefont {Jo}}, \bibinfo {author} {\bibfnamefont
  {Y.-I.}\ \bibnamefont {Shin}}, \bibinfo {author} {\bibfnamefont
  {W.}~\bibnamefont {Ketterle}}, \ and\ \bibinfo {author} {\bibfnamefont
  {D.}~\bibnamefont {Pritchard}},\ }\href@noop {} {\bibfield  {journal}
  {\bibinfo  {journal} {Phys. Rev. A}\ }\textbf {\bibinfo {volume} {78}},\
  \bibinfo {pages} {033429} (\bibinfo {year} {2008})}\BibitemShut {NoStop}%
\bibitem [{\citenamefont {Bajcsy}\ \emph {et~al.}(2009)\citenamefont {Bajcsy},
  \citenamefont {Hofferberth}, \citenamefont {Balic}, \citenamefont {Peyronel},
  \citenamefont {Hafezi}, \citenamefont {Zibrov}, \citenamefont {Vuletic},\
  and\ \citenamefont {Lukin}}]{bajcsy2009efficient}%
  \BibitemOpen
  \bibfield  {author} {\bibinfo {author} {\bibfnamefont {M.}~\bibnamefont
  {Bajcsy}}, \bibinfo {author} {\bibfnamefont {S.}~\bibnamefont {Hofferberth}},
  \bibinfo {author} {\bibfnamefont {V.}~\bibnamefont {Balic}}, \bibinfo
  {author} {\bibfnamefont {T.}~\bibnamefont {Peyronel}}, \bibinfo {author}
  {\bibfnamefont {M.}~\bibnamefont {Hafezi}}, \bibinfo {author} {\bibfnamefont
  {A.~S.}\ \bibnamefont {Zibrov}}, \bibinfo {author} {\bibfnamefont
  {V.}~\bibnamefont {Vuletic}}, \ and\ \bibinfo {author} {\bibfnamefont
  {M.~D.}\ \bibnamefont {Lukin}},\ }\href@noop {} {\bibfield  {journal}
  {\bibinfo  {journal} {Phys. Rev. Lett.}\ }\textbf {\bibinfo {volume} {102}},\
  \bibinfo {pages} {203902} (\bibinfo {year} {2009})}\BibitemShut {NoStop}%
\bibitem [{\citenamefont {Vetsch}\ \emph {et~al.}(2010)\citenamefont {Vetsch},
  \citenamefont {Reitz}, \citenamefont {Sagu{\'e}}, \citenamefont {Schmidt},
  \citenamefont {Dawkins},\ and\ \citenamefont
  {Rauschenbeutel}}]{vetsch2010optical}%
  \BibitemOpen
  \bibfield  {author} {\bibinfo {author} {\bibfnamefont {E.}~\bibnamefont
  {Vetsch}}, \bibinfo {author} {\bibfnamefont {D.}~\bibnamefont {Reitz}},
  \bibinfo {author} {\bibfnamefont {G.}~\bibnamefont {Sagu{\'e}}}, \bibinfo
  {author} {\bibfnamefont {R.}~\bibnamefont {Schmidt}}, \bibinfo {author}
  {\bibfnamefont {S.}~\bibnamefont {Dawkins}}, \ and\ \bibinfo {author}
  {\bibfnamefont {A.}~\bibnamefont {Rauschenbeutel}},\ }\href@noop {}
  {\bibfield  {journal} {\bibinfo  {journal} {Phys. Rev. Lett.}\ }\textbf
  {\bibinfo {volume} {104}},\ \bibinfo {pages} {203603} (\bibinfo {year}
  {2010})}\BibitemShut {NoStop}%
\bibitem [{\citenamefont {Shahmoon}\ \emph {et~al.}(2011)\citenamefont
  {Shahmoon}, \citenamefont {Kurizki}, \citenamefont {Fleischhauer},\ and\
  \citenamefont {Petrosyan}}]{shahmoon2011strongly}%
  \BibitemOpen
  \bibfield  {author} {\bibinfo {author} {\bibfnamefont {E.}~\bibnamefont
  {Shahmoon}}, \bibinfo {author} {\bibfnamefont {G.}~\bibnamefont {Kurizki}},
  \bibinfo {author} {\bibfnamefont {M.}~\bibnamefont {Fleischhauer}}, \ and\
  \bibinfo {author} {\bibfnamefont {D.}~\bibnamefont {Petrosyan}},\ }\href@noop
  {} {\bibfield  {journal} {\bibinfo  {journal} {Phys. Rev. A}\ }\textbf
  {\bibinfo {volume} {83}},\ \bibinfo {pages} {033806} (\bibinfo {year}
  {2011})}\BibitemShut {NoStop}%
\bibitem [{\citenamefont {Langbecker}\ \emph {et~al.}(2017)\citenamefont
  {Langbecker}, \citenamefont {Noaman}, \citenamefont {Kj{\ae}rgaard},
  \citenamefont {Benabid},\ and\ \citenamefont
  {Windpassinger}}]{langbecker2017rydberg}%
  \BibitemOpen
  \bibfield  {author} {\bibinfo {author} {\bibfnamefont {M.}~\bibnamefont
  {Langbecker}}, \bibinfo {author} {\bibfnamefont {M.}~\bibnamefont {Noaman}},
  \bibinfo {author} {\bibfnamefont {N.}~\bibnamefont {Kj{\ae}rgaard}}, \bibinfo
  {author} {\bibfnamefont {F.}~\bibnamefont {Benabid}}, \ and\ \bibinfo
  {author} {\bibfnamefont {P.}~\bibnamefont {Windpassinger}},\ }\href@noop {}
  {\bibfield  {journal} {\bibinfo  {journal} {Phys. Rev. A}\ }\textbf {\bibinfo
  {volume} {96}},\ \bibinfo {pages} {041402} (\bibinfo {year}
  {2017})}\BibitemShut {NoStop}%
\bibitem [{\citenamefont {He}\ \emph {et~al.}(2014)\citenamefont {He},
  \citenamefont {Sharypov}, \citenamefont {Sheng}, \citenamefont {Simon},\ and\
  \citenamefont {Xiao}}]{he2014two}%
  \BibitemOpen
  \bibfield  {author} {\bibinfo {author} {\bibfnamefont {B.}~\bibnamefont
  {He}}, \bibinfo {author} {\bibfnamefont {A.}~\bibnamefont {Sharypov}},
  \bibinfo {author} {\bibfnamefont {J.}~\bibnamefont {Sheng}}, \bibinfo
  {author} {\bibfnamefont {C.}~\bibnamefont {Simon}}, \ and\ \bibinfo {author}
  {\bibfnamefont {M.}~\bibnamefont {Xiao}},\ }\href@noop {} {\bibfield
  {journal} {\bibinfo  {journal} {Phys. Rev. Lett.}\ }\textbf {\bibinfo
  {volume} {112}},\ \bibinfo {pages} {133606} (\bibinfo {year}
  {2014})}\BibitemShut {NoStop}%
\bibitem [{\citenamefont {Johnson}\ and\ \citenamefont
  {Rolston}(2010)}]{johnson2010interactions}%
  \BibitemOpen
  \bibfield  {author} {\bibinfo {author} {\bibfnamefont {J.}~\bibnamefont
  {Johnson}}\ and\ \bibinfo {author} {\bibfnamefont {S.}~\bibnamefont
  {Rolston}},\ }\href@noop {} {\bibfield  {journal} {\bibinfo  {journal} {Phys.
  Rev. A}\ }\textbf {\bibinfo {volume} {82}},\ \bibinfo {pages} {033412}
  (\bibinfo {year} {2010})}\BibitemShut {NoStop}%
\bibitem [{\citenamefont {Jau}\ \emph {et~al.}(2015)\citenamefont {Jau},
  \citenamefont {Hankin}, \citenamefont {Keating}, \citenamefont {Deutsch},\
  and\ \citenamefont {Biedermann}}]{jau2015entangling}%
  \BibitemOpen
  \bibfield  {author} {\bibinfo {author} {\bibfnamefont {Y.-Y.}\ \bibnamefont
  {Jau}}, \bibinfo {author} {\bibfnamefont {A.}~\bibnamefont {Hankin}},
  \bibinfo {author} {\bibfnamefont {T.}~\bibnamefont {Keating}}, \bibinfo
  {author} {\bibfnamefont {I.}~\bibnamefont {Deutsch}}, \ and\ \bibinfo
  {author} {\bibfnamefont {G.}~\bibnamefont {Biedermann}},\ }\href@noop {}
  {\bibfield  {journal} {\bibinfo  {journal} {Nat. Phys.}\ }\textbf {\bibinfo
  {volume} {12}} (\bibinfo {year} {2015})}\BibitemShut {NoStop}%
\bibitem [{\citenamefont {Li}\ and\ \citenamefont
  {Kuzmich}(2016)}]{Li2016quantum}%
  \BibitemOpen
  \bibfield  {author} {\bibinfo {author} {\bibfnamefont {L.}~\bibnamefont
  {Li}}\ and\ \bibinfo {author} {\bibfnamefont {A.}~\bibnamefont {Kuzmich}},\
  }\href@noop {} {\bibfield  {journal} {\bibinfo  {journal} {Nat. Commun.}\
  }\textbf {\bibinfo {volume} {7}},\ \bibinfo {pages} {13618} (\bibinfo {year}
  {2016})}\BibitemShut {NoStop}%
\bibitem [{\citenamefont {Yu}\ and\ \citenamefont
  {Fan}(2009)}]{yu2009complete}%
  \BibitemOpen
  \bibfield  {author} {\bibinfo {author} {\bibfnamefont {Z.}~\bibnamefont
  {Yu}}\ and\ \bibinfo {author} {\bibfnamefont {S.}~\bibnamefont {Fan}},\
  }\href@noop {} {\bibfield  {journal} {\bibinfo  {journal} {Nat. Photon.}\
  }\textbf {\bibinfo {volume} {3}},\ \bibinfo {pages} {91} (\bibinfo {year}
  {2009})}\BibitemShut {NoStop}%
\bibitem [{\citenamefont {Mrejen}\ \emph {et~al.}(2015)\citenamefont {Mrejen},
  \citenamefont {Suchowski}, \citenamefont {Hatakeyama}, \citenamefont {Wu},
  \citenamefont {Feng}, \citenamefont {O¡¯Brien}, \citenamefont {Wang},\ and\
  \citenamefont {Zhang}}]{mrejen2015adiabatic}%
  \BibitemOpen
  \bibfield  {author} {\bibinfo {author} {\bibfnamefont {M.}~\bibnamefont
  {Mrejen}}, \bibinfo {author} {\bibfnamefont {H.}~\bibnamefont {Suchowski}},
  \bibinfo {author} {\bibfnamefont {T.}~\bibnamefont {Hatakeyama}}, \bibinfo
  {author} {\bibfnamefont {C.}~\bibnamefont {Wu}}, \bibinfo {author}
  {\bibfnamefont {L.}~\bibnamefont {Feng}}, \bibinfo {author} {\bibfnamefont
  {K.}~\bibnamefont {O¡¯Brien}}, \bibinfo {author} {\bibfnamefont
  {Y.}~\bibnamefont {Wang}}, \ and\ \bibinfo {author} {\bibfnamefont
  {X.}~\bibnamefont {Zhang}},\ }\href@noop {} {\bibfield  {journal} {\bibinfo
  {journal} {Nat. Commun.}\ }\textbf {\bibinfo {volume} {6}} (\bibinfo {year}
  {2015})}\BibitemShut {NoStop}%
\end{thebibliography}%

\end{document}